\pdfoutput=1
\documentclass[11pt,twoside,a4paper,cmspaper,final,collab]{cms-tdr}
\def\svnVersion{1e11d29}\def\svnDate{2023/12/08}\def\cmsCernNoTag{CERN-EP-2023-115}\def\cmsCernDate{\today}\def\cmsMessage{Published in the Journal of High Energy Physics as \href{http://dx.doi.org/10.1007/JHEP11(2023)181}{\doi{10.1007/JHEP11(2023)181}.}}
\begin{document}\cmsNoteHeader{EXO-20-006}

\cmsNoteHeader{EXO-20-006}

\DeclareRobustCommand{\N}{{\HepParticle{\PN}{\ell}{}}\xspace}
\DeclareRobustCommand{\Ne}{{\HepParticle{\PN}{\Pe}{}}\xspace}
\DeclareRobustCommand{\Nm}{{\HepParticle{\PN}{\PGm}{}}\xspace}
\DeclareRobustCommand{\Nt}{{\HepParticle{\PN}{\PGt}{}}\xspace}
\DeclareRobustCommand{\WRpm}{{\HepParticle{\PW}{R}{\pm}}\xspace}

\newcommand{\mN}{\ensuremath{m_{\N}}\xspace}
\newcommand{\mZp}{\ensuremath{m_\PZpr}\xspace}
\newcommand{\SU}{\ensuremath{\mathrm{SU}}\xspace}
\newcommand{\U}{\ensuremath{\mathrm{U}}\xspace}

\providecommand{\cmsTable}[1]{\resizebox{\textwidth}{!}{#1}}

\title{Search for \texorpdfstring{\PZpr}{Z'} bosons decaying to pairs of heavy Majorana neutrinos in proton-proton collisions at \texorpdfstring{$\sqrt{s} = 13\TeV$}{sqrt(s) = 13 TeV}}

\date{\today}

\abstract{
A search for the production of pairs of heavy Majorana neutrinos (\N) from the decays of \PZpr bosons is performed using the CMS detector at the LHC. The data were collected in proton-proton collisions at a center-of-mass energy of $\sqrt{s} = 13\TeV$, with an integrated luminosity of 138\fbinv. The signature for the search is an excess in the invariant mass distribution of the final-state objects, two same-flavor leptons (\Pe or \Pgm) and at least two jets. No significant excess of events beyond the expected background is observed. Upper limits at 95\% confidence level are set on the product of the \PZpr production cross section and its branching fraction to a pair of \N, as functions of \N and \PZpr boson masses (\mN and \mZp, respectively) for \mZp from 0.4 to 4.6\TeV and \mN from 0.1\TeV to $\mZp/2$. In the theoretical framework of a left-right symmetric model, exclusion bounds in the $\mN$-$\mZp$ plane are presented in both the electron and muon channels. The observed upper limit on \mZp reaches up to 4.42\TeV. These are the most restrictive limits to date on the mass of \N as a function of the \PZpr boson mass.
}

\hypersetup{
pdfauthor={CMS Collaboration},
pdftitle={Search for Z' bosons decaying to pairs of heavy Majorana neutrinos in proton-proton collisions at sqrt(s)=13 TeV},
pdfsubject={CMS},
pdfkeywords={CMS, beyond standard model, Majorana neutrino}}

\maketitle

\section{Introduction}
The observation of neutrino oscillations~\cite{pdg,SNO:2002tuh,Super-Kamiokande:1998kpq,Super-Kamiokande:2004orf}
established that at least two of the standard model (SM) neutrinos have mass.
The nonzero masses of the neutrinos are clear evidence of physics beyond the SM, in which neutrinos are massless.
Upper limits on the neutrino masses have been obtained from cosmological observations~\cite{PhysRevD.104.083504}.
A direct measurement from tritium beta decays~\cite{KATRIN} indicates that the electron-type neutrino mass is less than 0.8\unit{eV}.
The fact that neutrino masses are so much smaller than the other fermion masses and that right-handed neutrinos are not observed suggest that neutrino masses may have an origin other than a Yukawa coupling to the Higgs field.

{\tolerance=1200
In addition to its inability to explain neutrino mass, the SM does not provide any clear answer as to what is the source of the parity violation in the weak sector or of the matter dominance of the universe.
One of the leading theoretical solutions is to introduce a left-right symmetry model (LRSM)~\cite{Maiezza:2010ic}.
The LRSM is based on a gauge group of ${\SU(3)}_\mathrm{C} \bigotimes {\SU(2)}_\mathrm{L} \bigotimes {\SU(2)}_\mathrm{R} \bigotimes {\U(1)}_\mathrm{B-L}$, established by a symmetry between the left- and right-handed $\SU(2)$ groups.
In this model, there are three additional gauge bosons, \WRpm and \PZpr, as well as three right-handed neutrinos (\Ne, \Nm, and \Nt).
The spontaneous symmetry breaking of the LRSM, at some energy scale, can generate the gauge group of the SM in a way that naturally includes the observed parity violation in the weak sector, and also provides an additional source of CP violation that can explain the asymmetry between matter and antimatter in the universe.
Heavy right-handed neutrinos and light left-handed neutrinos are also naturally generated by the symmetry breaking, through a process referred to as the see-saw mechanism~\cite{seesaw1,seesaw2,seesaw3,PhysRevLett.44.912,DOI1981323}.
\par}

\begin{figure}[htbp]
\centering
\includegraphics[width=0.8\textwidth]{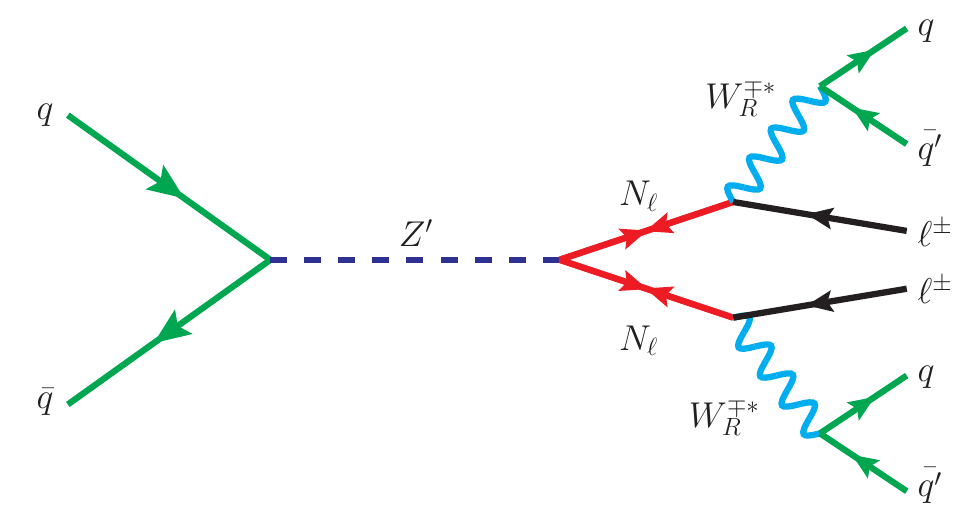}
\caption{Feynman diagram representing the pair production of \N via \PZpr boson exchange, where $\ell = \Pe$ or \Pgm. Each \N decays into a lepton and two quarks, and we assume that the \WRpm is heavier than the \N. Both opposite- and same-sign dileptons from the decays of the heavy neutrino pair are allowed, because of the Majorana nature of \N.}
\label{fig:HN_pair_feynman}
\end{figure}

As the LRSM provides explanations for these basic observations of beyond the SM physics, it is compelling to probe signatures of this model.
Although there have been previous searches for both new heavy gauge bosons and heavy right-handed neutrinos~\cite{CMS_Zp_dilep,CMS:2022yjm,CMS_HNL_dilep,PhysRevLett.120.221801,CMS:2022fut},
the unique signature of the LRSM is the presence of events
that have an extra gauge boson as well as heavy right-handed neutrinos. In this analysis, we search for the pair production of heavy right-handed neutrinos through an extra neutral gauge boson (\PZpr).
The resonance production of a \PZpr boson and its decay into heavy neutrinos are shown in Fig.~\ref{fig:HN_pair_feynman}.
In the benchmark model used in this study, the heavy right-handed neutrino is a Majorana particle and decays into a lepton plus two quarks.
We therefore select events with two electrons or two muons, and at least two reconstructed jets, to search for an excess in the invariant mass distribution of them.
We use opposite- and same-sign leptons inclusively, and mixing between heavy neutrinos of electron type and muon type is not considered.
Rich scalar sectors are predicted by LRSM models in general. In this search, results are presented in terms of the product of the cross section and branching fraction of the process,
thus no assumptions are made concerning the nature of the scalar sector.
Detailed underlying assumptions for this LRSM will be discussed in Section~\ref{sec:Samples}.

Searches for a \WRpm boson and its decay to a heavy Majorana neutrino (\N) have been performed in the framework of the LRSM by the ATLAS and the CMS Collaborations at the LHC~\cite{ATLAS:2012ak,CMS:2012zv,Khachatryan:2014dka,Aad:2015xaa,WR_CMS_2015,WR_CMS_2016_ICHEP,Sirunyan:2018pom,Aaboud:2018spl,Aaboud:2019wfg,CMS:2021dzb,ATLAS:2023cjo}.
The most stringent lower limit on the \WRpm mass from the ATLAS Collaboration is about 6.4\TeV~\cite{ATLAS:2023cjo},
which was obtained using a data set of proton-proton ($\Pp\Pp$) collisions at a center-of-mass energy of 13\TeV, corresponding to an integrated luminosity of 139\fbinv.
A search for an extra gauge boson \PZpr and its decay to a pair of \N has also been performed by the ATLAS Collaboration,
which obtained a lower limit on the \PZpr mass (\mZp) of 2.2\TeV~\cite{Aad:2015xaa} using a data sample corresponding to 20.3\fbinv at $\sqrt{s} = 8\TeV$.

All these results are limited by a reduced efficiency for the signals in the kinematic region where the heavy neutrino mass (\mN) is much smaller than the mass of the extra gauge boson.
A lepton from the decay of a highly Lorentz-boosted heavy neutrino overlaps with jets from the same decay, and lepton identification requires the use of isolation criteria. As a result, the signal selection efficiency is reduced in this kinematic region.
It is important to recover the efficiency in this kinematic region, because the dominant decay of extra gauge bosons is to heavy neutrinos.
Here we present a method that is sensitive to a broad range of \N masses, including the region where \mN is much smaller than \mZp.
The analysis presented here uses a $\Pp\Pp$ collision data set at $\sqrt{s} = 13\TeV$ collected with the CMS detector, corresponding to an integrated luminosity of 138\fbinv.

Tabulated results are provided in the HEPData record for this analysis~\cite{hepdata}.

\section{The CMS detector}
\label{sec:CMS}
The central feature of the CMS apparatus is a superconducting solenoid of 6\unit{m} internal diameter, providing a magnetic field of 3.8\unit{T}. Within the solenoid volume are a silicon pixel and strip tracker,
a lead tungstate crystal electromagnetic calorimeter (ECAL), and a brass and scintillator hadron calorimeter (HCAL), each composed of a barrel and two endcap sections. Forward calorimeters extend the pseudorapidity ($\eta$) coverage provided
by the barrel and endcap detectors. Muons are detected in gas-ionization detectors embedded in the steel flux-return yoke outside the solenoid.
A more detailed description of the CMS detector, together with a definition of the coordinate system used and the relevant kinematic variables, can be found in Ref.~\cite{Chatrchyan:2008zzk}.

\section{Simulated samples}
\label{sec:Samples}
Signal and background events for each data-taking year from 2016 to 2018 are produced using Monte Carlo (MC) event generators and a detailed simulation of the CMS detector.
The expected background processes include Drell--Yan (DY) dilepton plus jets production, and production of \ttbar, dibosons, $\PW+$jets, single top quarks, and QCD multijet events.
The DY events are simulated with \MGvATNLO2.2.2~\cite{Alwall:2014hca}
with up to four jets at leading order (LO) in quantum chromodynamics (QCD) for 2016. For 2017 and 2018, \MGvATNLO2.4.2 is used.
The \PW events associated with jets are obtained with the \MGvATNLO2.3.3~\cite{Alwall:2014hca} generator with up to four jets at LO.
The MLM scheme~\cite{Alwall:2007fs} is used to match partons from matrix element calculations and parton showers for these two samples.
The \ttbar production is simulated with the \POWHEG v2~\cite{Nason:2004rx,Frixione:2007vw,Alioli:2010xd,Frixione:2007nw} generator at next-to-LO (NLO).
Diboson ($\PW\PW$, $\PW\PZ$, and $\PZ\PZ$) events are simulated with the \PYTHIA 8.212~\cite{Sjostrand:2014zea} generator at LO.
The $t$-channel and $\PQt\PW$ single top quark events are produced with the \POWHEG v1~\cite{Nason:2004rx,Frixione:2007vw,Alioli:2010xd,Alioli:2009je,Re:2010bp} generator.
Single top quark production in the $s$-channel is simulated with the \MGvATNLO generator at NLO precision.
The \ttbar production in association with a \PW (\PZ) boson is simulated using \MGvATNLO at NLO (LO) precision.
QCD multijet events are produced using \MGvATNLO at LO precision.

The NNPDF3.0~\cite{Ball:2014uwa} parton distribution function (PDF) set is used to simulate background events corresponding to the data recorded in 2016.
For the 2017 and 2018 backgrounds and all signal events, the NNPDF3.1~\cite{Ball:2017nwa} PDF set is used.
Parton showering, including photon radiation and hadronization processes, with the choice of NNPDF3.0 (or 3.1) are modeled using \PYTHIA 8.226 (8.230)~\cite{Sjostrand:2014zea}
with the CUETP8M1 (CP5)~\cite{Khachatryan:2015pea,Sirunyan:2019dfx} tune.
Finally, the response of the CMS detector is modeled using \GEANTfour~\cite{Geant4} for all simulated samples.

The signal events for the process shown in Fig.~\ref{fig:HN_pair_feynman} are simulated using \MGvATNLO2.6.0 at NLO precision with the LRSM model cards~\cite{Mattelaer:2016ynf}.
Heavy neutrinos with electron and muon flavors are considered, and no mixing between the two flavors is assumed.
Both opposite- and same-sign configurations of dileptons are taken into account in the simulation.
The signal samples are simulated for various \mZp and \mN hypotheses, with \mZp between 400 and 5000\GeV, and \mN between 100\GeV and $\mZp/2$.
In the simulation, the \PWR mass is assumed to be 5\TeV, so that the heavy neutrinos are always lighter than the \PWR boson.
The product of the NLO cross section and the branching fraction (BF) is the same for both dilepton channels and is 27.77\unit{pb} for $\mZp=400\GeV$, and $8.60 \times 10^{-5}\unit{pb}$ for $\mZp=5000\GeV$ with $\mN= 100\GeV$. With $\mN= 2400\GeV$,
it is $4.94 \times 10^{-7}\unit{pb}$ for $\mZp=5000\GeV$.
A product of cross section and branching fraction of 1\unit{fb} corresponds typically to the \PZpr boson masses in the range 3 to 4\TeV, depending slightly on \mN.

Minimum bias events generated with \PYTHIA are superimposed on each simulated hard scattering event to reproduce the effect of extra $\Pp\Pp$ interactions within the same or neighboring bunch crossings (pileup).
The simulated events are weighted such that the distribution of the number of additional pileup interactions, estimated from the measured instantaneous luminosity for each bunch crossing,
matches that observed in data sets individually for each year of data-taking in 2016--2018.
The simulated events are processed with the same reconstruction software used for $\Pp\Pp$ collision data.

\section{Event reconstruction and object identification}
\label{sec:EvtRecoID}
Candidate events from decays of heavy neutrino pairs are selected using a two-tiered trigger system~\cite{Khachatryan:2016bia}.
The first level (L1), composed of custom hardware processors, uses information from the calorimeters and
muon detectors to select events at a rate of around 100\unit{kHz} within a fixed time interval of less than 4\mus. The second level, known as the high-level trigger (HLT), consists of a farm of processors
running a version of the full event reconstruction software optimized for fast processing, and reduces the event rate to around 1\unit{kHz} before data storage.

For dielectron channel events, a diphoton trigger is used, which requires two electron or photon ($\Pe/\gamma$) objects detected in the ECAL with a minimum transverse energy (\ET) of
60\GeV for 2016 data or 70\GeV for 2017 and 2018 data.
Dimuon channel events must pass a single-muon trigger that requires the presence of a muon with a minimum \pt of 50\GeV.

In the dielectron channel, the trigger efficiency of signal events is about 85\%, dropping to about 10\% in the kinematic region
where \mN is much smaller than \mZp.
The reduction of efficiency is mainly caused by the requirement on H/E, the ratio between energies deposited in HCAL and ECAL, being smaller than 0.15 (barrel) or 0.10 (endcap).
Since the requirement on H/E is essential for distinguishing electrons and photons from hadrons,
there are inevitable signal losses in the dielectron channel for highly Lorentz-boosted events.
This analysis uses a double-photon trigger, since it has the loosest H/E requirement among all $\Pe/\gamma$ HLT paths.

The trigger efficiency for signal events in the dimuon channel depends on \mN and \mZp, and ranges between 80 and 95\%, except at the lightest \PZpr boson mass points,
where it is somewhat lower because of the high momentum threshold of the muon trigger.
For example an efficiency of about 60\% is observed for the mass points at the 400\GeV and 600\GeV.
The cross sections for these mass points are higher than those for heavier \PZpr bosons.
This allows good sensitivity to be maintained while keeping the same trigger for all masses.

The primary vertex (PV) is taken to be the vertex corresponding to the hardest scattering in the event, evaluated using tracking information alone,
as described in Section 9.4.1 of Ref.~\cite{CMS-TDR-15-02}.

Particles in selected events are reconstructed and identified using the global event reconstruction
(also called particle-flow event reconstruction~\cite{CMS-PRF-14-001}) with an optimized combination of all subdetector information.
The identification of the particle type (photon, electron, muon, charged hadron, neutral hadron) plays an important role in the determination of the particle direction and energy. Photons (\eg, coming from \Pgpz\ decays or from electron bremsstrahlung)
are identified as ECAL energy clusters not linked to the extrapolation of any charged particle trajectory to the ECAL.
Electrons (\eg, from \PZ boson leptonic decays) are identified
as a primary charged particle track and potentially many ECAL energy clusters corresponding to this track extrapolation to the ECAL and to possible bremsstrahlung photons emitted while traversing the tracker material.
Muons (\eg, from \PW boson leptonic decays) are identified as tracks in the central tracker matching with either a track or several hits in the muon system, and associated with calorimeter deposits compatible with the muon hypothesis.
Charged hadrons are identified as charged particle tracks neither identified as electrons, nor as muons. Finally, neutral hadrons are identified as HCAL energy clusters not linked to any charged hadron trajectory,
or as a combined ECAL and HCAL energy excess with respect to the expected charged hadron energy deposit.

The energy of photons is obtained from the ECAL measurement. The energy of electrons is determined from a combination of the track momentum at the PV, the corresponding ECAL cluster energy,
and the energy sum of all bremsstrahlung photons associated with the track.
The energy of muons is obtained from the corresponding track momentum. The energy of charged hadrons is determined from a combination of
the track momentum and the corresponding ECAL and HCAL energies, corrected for the response function of the calorimeters to hadronic showers.
Finally, the energy of neutral hadrons is obtained from the corresponding corrected ECAL and HCAL energies.

\subsection{Lepton selection}
Electrons are selected in the region of $\abs{{\eta}_\mathrm{C}} < 2.5$, where ${\eta}_\mathrm{C}$ is the pseudorapidity of the ECAL cluster with respect to the nominal center of the CMS detector.
The transition region between ECAL barrel and endcaps, $1.44 < \abs{{\eta}_\mathrm{C}} < 1.57$, is excluded. 
The electron \ET is required to be greater than 65 (75)\GeV for the 2016 (2017 and 2018) data.
In this analysis, two electron identification criteria are used: ``loose'' and ``tight''.
A ``loose'' electron is an object passing the ``HEEP'' selection or passing the selection of a loose cut-based electron with no isolation criteria~\cite{CMS:2020uim,CMS-DP-2020-021}.
The ``HEEP'' criteria have been developed for the analysis of high-energy electrons.
A ``tight'' electron is an object passing the ``HEEP'' requirement.

In addition, a ``veto'' electron is defined using the same criteria as the loose electron but lowering the \ET requirement to 10\GeV.

Muons are selected in the region $\abs{\eta} < 2.4$ with $\pt > 65\,(75)\GeV$ for 2016 (2017 and 2018). 
Higher \pt requirements for muons compared to the trigger thresholds are applied to reduce QCD background.
A dedicated muon identification algorithm, HighPtID~\cite{Sirunyan_2018}, adapted from searches for resonances using high-momentum leptons~\cite{Sirunyan:2018exx} is used for this analysis.
Two muon identification criteria are used, ``loose'' and ``tight''.
A ``loose'' muon is required to satisfy the HighPtID requirements. A ``tight'' muon must pass both the HighPtID and isolation requirements.
The isolation is defined as the \pt sum of tracks within a cone of radius $\Delta R=\sqrt{\smash[b]{(\Delta\eta)^2+(\Delta\phi)^2}} = 0.3$ around the muon candidate direction, where $\phi$ is its azimuthal angle in radians.
The momentum of the muon candidate is excluded from the sum. If the sum is less than 10\% of the muon candidate's \pt, it passes the isolation requirement.
In addition, a ``veto'' muon is defined using the particle-flow muons with $\pt > 10\GeV$.

\subsection{Jet selection}
For each event, hadronic jets are reconstructed by clustering particle-flow objects using the anti-\kt algorithm~\cite{Cacciari:2008gp, Cacciari:2011ma} with two distance
parameters of 0.4 and 0.8, producing AK4 and AK8 jets, respectively.
The jet momentum is determined as the vectorial sum of all particle momenta in the jet, and is found from the simulation to be within 5 to 10\% of the true momentum, on average, over the entire \pt spectrum and detector
acceptance. Pileup interactions can contribute extra tracks and calorimetric energy depositions to the jet momentum.

In order to mitigate the effect of pileup, the pileup-per-particle identification algorithm~\cite{Bertolini:2014bba, CMS:2020ebo} is used at the reconstructed-particle level, making use of local shape information,
event pileup properties and tracking information. Charged particles identified to be originating from pileup vertices are discarded. For each neutral particle, a local shape variable is computed using
the surrounding charged particles compatible with the PV within the tracker acceptance ($\abs{\eta} < 2.5$), and using both charged and neutral particles in the region outside of the tracker coverage.
The momenta of the neutral particles are then rescaled according to their probability to originate from the PV deduced from the local shape variables, superseding the need for jet-based pileup corrections~\cite{CMS-PAS-JME-16-003}.
For this analysis, AK4 (AK8) jets are required to have $\pt > 40\,(300)\GeV$ and $\abs{\eta} < 2.7$.

AK8 jets are groomed using the soft-drop algorithm~\cite{Dasgupta:2013ihk,Butterworth:2008iy}. In this algorithm, the constituents of the AK8 jets are reclustered using the Cambridge--Aachen algorithm~\cite{Dokshitzer:1997in,Wobisch:1998wt}. Soft radiation and wide-angle radiation from the jet are removed by setting the angular exponent $\beta$ to zero, the soft cutoff threshold to 0.1, and the characteristic radius $R_{0}$ to 0.8~\cite{Larkoski:2014wba}.
All AK8 jets are required to have mass greater than 40\GeV to reduce the number of AK8 jets that originate from a single parton.

In order to avoid double counting of AK4 jets with a lepton or an AK8 jet, any AK4 jet located within $\Delta R < 0.4$ of a loose lepton or
within $\Delta R < 1.0$ of an AK8 jet is not used in this analysis.

\section{Event selections}
\label{sec:EvtSel}

The signature targeted in this search is two same-flavor leptons and four jets from heavy neutrino decays, as shown in Fig.~\ref{fig:HN_pair_feynman}.
The kinematic distributions of the final state are strongly dependent on the ratio of twice the heavy neutrino mass and the \PZpr boson mass ($2\mN/\mZp$).
If the ratio is much smaller than one, the lepton and two jets from a heavy neutrino decay are merged into one jet because of the high Lorentz boost provided by the heavy neutrino.
As a consequence, leptons are mostly reconstructed as loose leptons since  they seldom satisfy the isolation criteria.
In contrast, leptons and jets are well separated if the ratio is close to unity.
A dedicated strategy is developed to achieve a good signal selection efficiency in the wide kinematic region covered by this search.

The selected events are required to have exactly two same-flavor ($\Pe\Pe$ or $\Pgm\Pgm$) loose leptons
using both the opposite- and same-sign configurations.
Events with additional veto leptons are removed.
To suppress the background contribution from DY events, the mass of the dilepton pair ($m_{\ell\ell}$) is required to be greater than 150\GeV.

No \PQb-tagging veto criteria are considered in the analysis, since high mass heavy neutrinos have a significant branching fraction into bottom quark jets, via $\N \rightarrow {\ell}^{-} \PQt \PAQb$ and $\N \rightarrow {\ell}^{+} \PAQt \PQb$ decays.
Furthermore, for lower heavy neutrino masses, final states typically contain jets with very high energies, for which \PQb-tagging performance is degraded and, as a result, systematic uncertainties prevent an improvement in sensitivity.

We define three different signal regions (SRs), depending on the number of AK8 jets found in the event, as shown in Table~\ref{tab:evt_sel_N_obj}.
The region SR1 contains no AK8 jets, SR2 contains exactly one AK8 jet, and SR3 contains at least two AK8 jets.

In SR1, we require two tight leptons and at least four AK4 jets, but no AK8 jets.
To reconstruct two heavy neutrinos for an event, the two tight leptons and the leading four AK4 jets are used.
Each heavy neutrino candidate is reconstructed using one lepton and two AK4 jets.
From the 12 possible lepton plus two-jet combination combinations, we select the one that minimizes the difference between the masses of the two heavy neutrino candidates.

In SR2, we require exactly one AK8 jet, at least one tight lepton and at least two AK4 jets.
In events with only one tight lepton, this lepton is paired with the two leading AK4 jets to reconstruct one heavy neutrino. The AK8 jet is assigned as a proxy to the other heavy neutrino.
If an event has two tight leptons, the AK8 jet is paired with the closer lepton, and the other lepton is paired with the two leading AK4 jets.

In SR3, we require at least two AK8 jets, and loose leptons are ignored.
In cases where an event has no tight leptons, the two leading AK8 jets are assigned as proxies to the two heavy neutrinos.
If there is only one tight lepton, the lepton is paired with the closer AK8 jet out of the two leading AK8 jets to reconstruct one heavy neutrino.
The remaining AK8 jet is assigned as a proxy to the other heavy neutrino.
If there are two tight leptons, the leading AK8 jet is paired to the closer lepton while the secondary AK8 jet and the other lepton are used to reconstruct the other heavy neutrino.

In all three SRs, the invariant mass of the \PZpr boson is obtained using the two reconstructed heavy neutrinos.
If a tight lepton is included in the energy of the AK8 jet, the four-momentum of the AK8 jet is assigned to the four-momentum of the heavy neutrino.
For $\mN = 100\GeV$ and $\mZp = 4\TeV$, about 2\% of the signal events that enter the SR3 are of this type.
Once the \PZpr boson and heavy neutrinos are reconstructed, additional requirements (reconstructed $\mN > 80\GeV$ and reconstructed $\mZp > 300\GeV$)
are applied to reduce backgrounds, and we start search range from $\mZp = 400\GeV$ and $\mN = 100\GeV$.

The products of acceptance and efficiency for the simulated signal samples are 20--25 (25--33)\% and 1.2--4.7 (7--16)\% in the dielectron (dimuon) channel for the most resolved and boosted mass points, respectively,
depending on \mZp.

\begin{table}[ptb]
\topcaption{Multiplicity requirements for AK8 jets, tight leptons and AK4 jets in the different signal regions considered in the search.}
\centering
\begin{tabular}{l c c c}
SR & $N$(AK8 jet) & $N$(tight leptons) & $N$(AK4 jet) \\
\hline
SR1 &  $=$0  & $=$2  & $\ge$4 \\
SR2 &  $=$1  & $\ge$1 & $\ge$2 \\
SR3 &  $\ge$2 & \NA & \NA \\
\end{tabular}
\label{tab:evt_sel_N_obj}
\end{table}

\section{Background estimation}
\label{sec:Bkg}
The backgrounds are dominated by SM processes containing prompt leptons. These are estimated using the simulated samples described in Section 3.
The leading background contribution is from \ttbar events.
The simulated \ttbar sample is normalized using the QCD next-to-NLO cross section,
which includes resummation of next-to-next-to-leading logarithmic soft gluon terms for the top quark mass of 172.5\GeV~\cite{Czakon:2011xx}.
A control region (CR), CR1, is defined to validate the \ttbar background using an $\Pe\Pgm$ sideband passing the same single-muon triggers used in the dimuon SR.
The region CR1 has exactly the same event selection as the SR, except the flavors of the two leptons are required to be different.
The events of CR1 are classified in three orthogonal regions according to the same criteria used for the SR, which are summarized in Table~\ref{tab:evt_sel_N_obj}.
Since no mixing between generations of heavy neutrinos is assumed and the heavy neutrino always decays into a lepton and two jets in the signal process, the signal in CR1 is negligible.
This expectation is supported by the observed consistency between simulated background predictions in CR1 and the corresponding data-driven estimates.
Additional scale factors are applied to the \ttbar simulated samples, depending on the AK8 jet multiplicity.
The scale factors are extracted from a simultaneous fit to all CRs and SRs and range from 0.83 to 1.06.
The uncertainties in these scale factors, determined from the fit, are assumed to cover the theoretical uncertainties in the simulated \ttbar distributions.

\begin{figure}[!hptb]
\centering
    \includegraphics[width=0.45\textwidth]{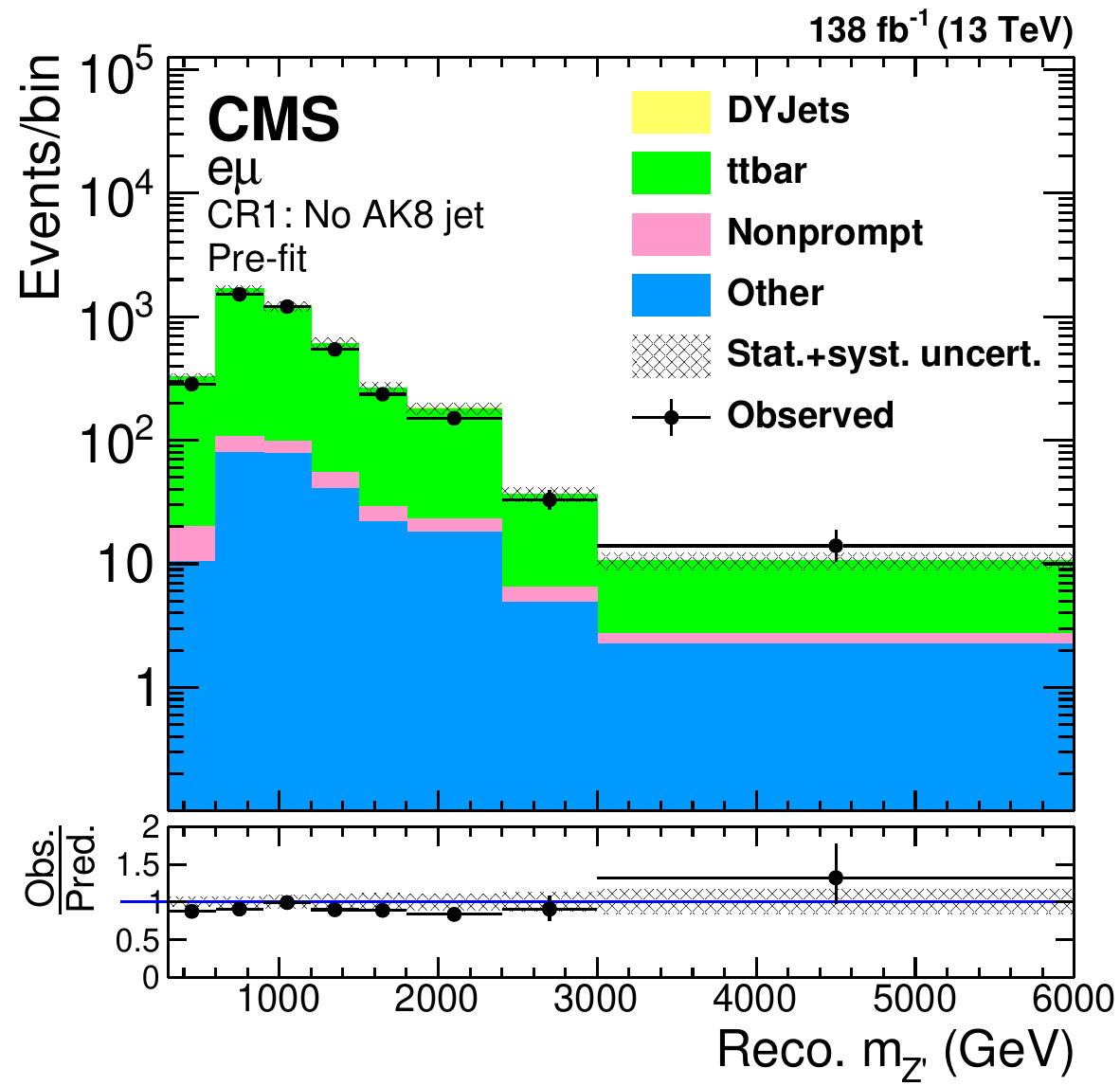}
    \hspace{0.01\textwidth}
    \includegraphics[width=0.45\textwidth]{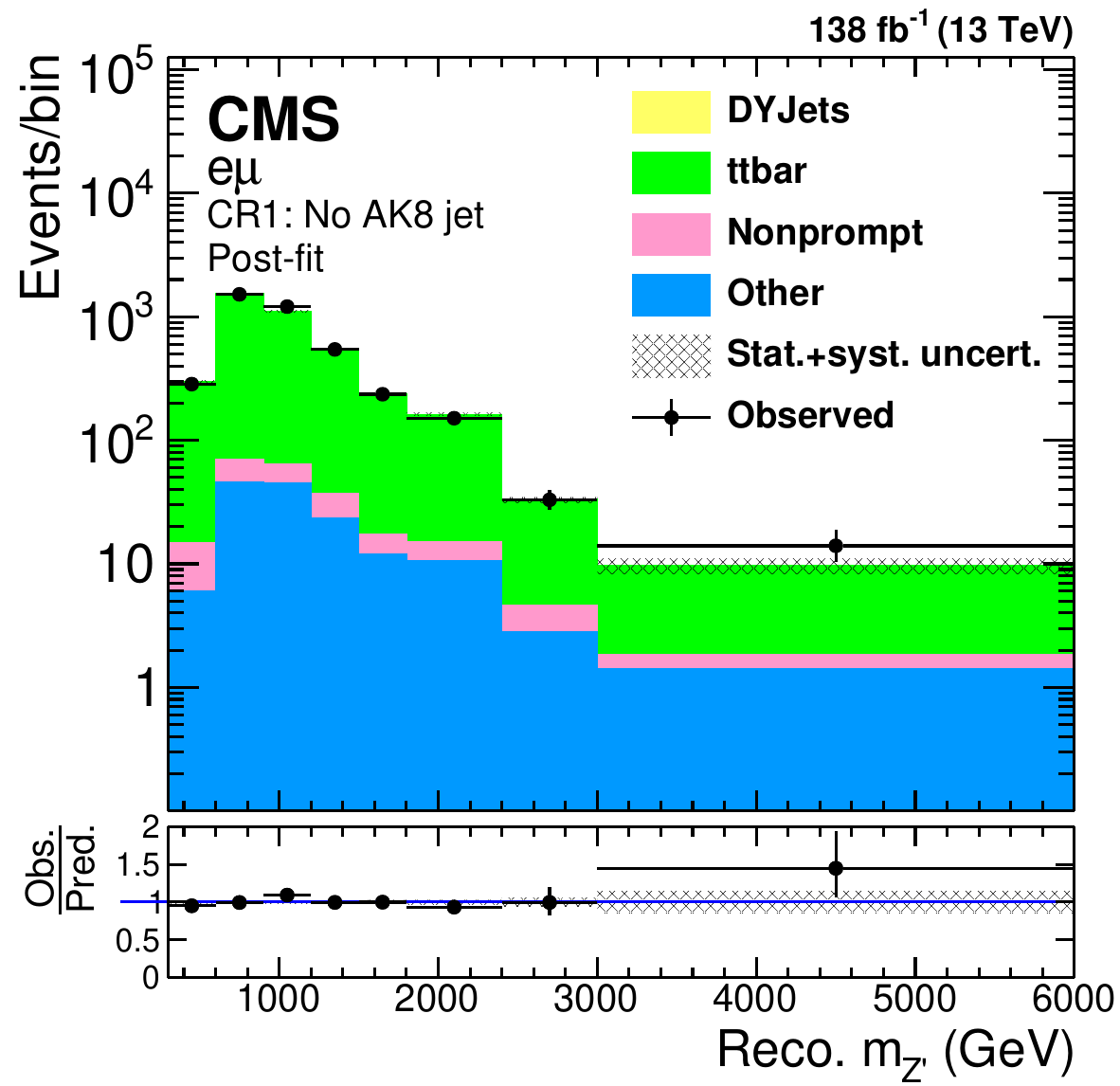}
    \hspace{0.01\textwidth}
    \includegraphics[width=0.45\textwidth]{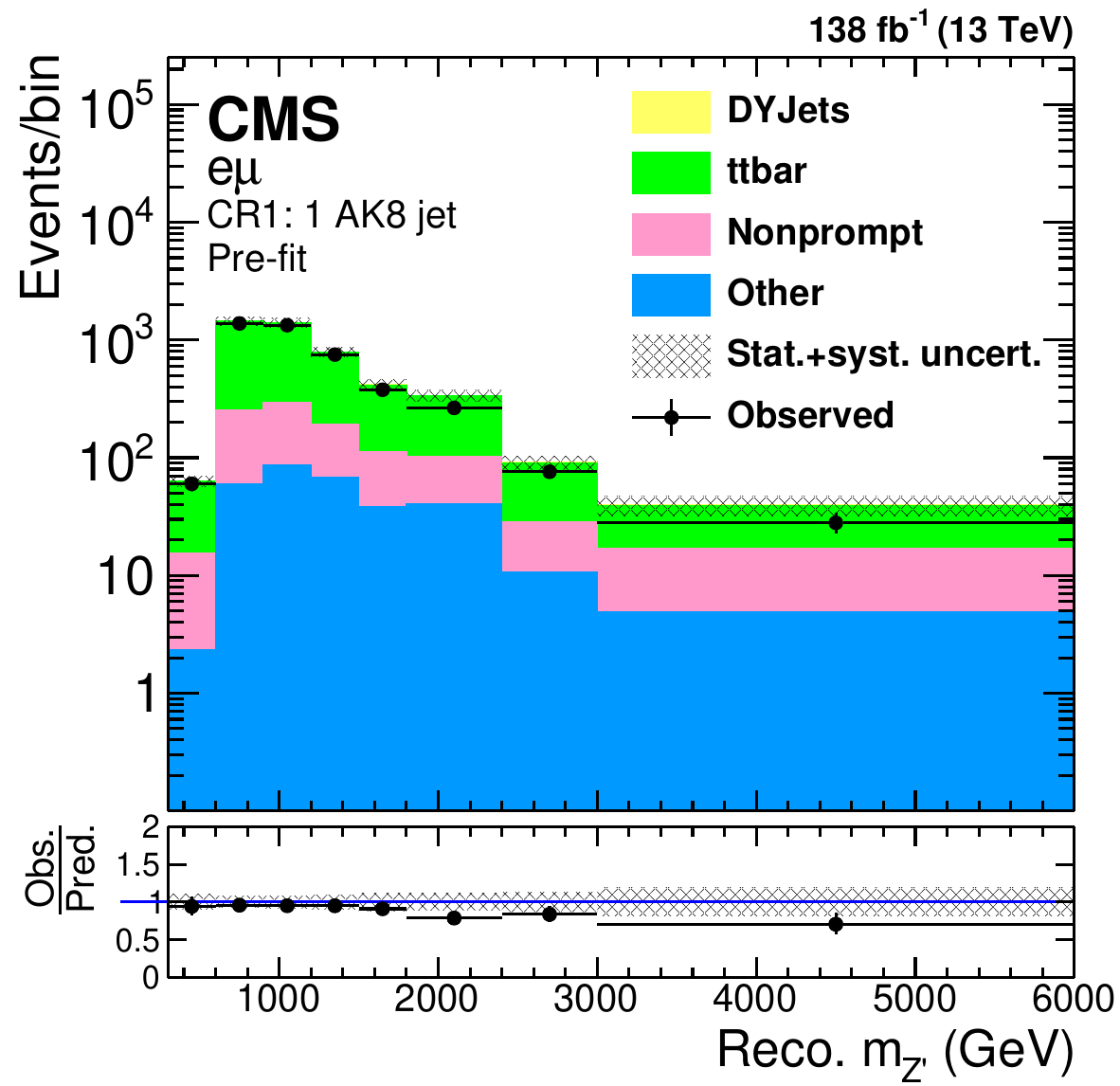}
    \hspace{0.01\textwidth}
    \includegraphics[width=0.45\textwidth]{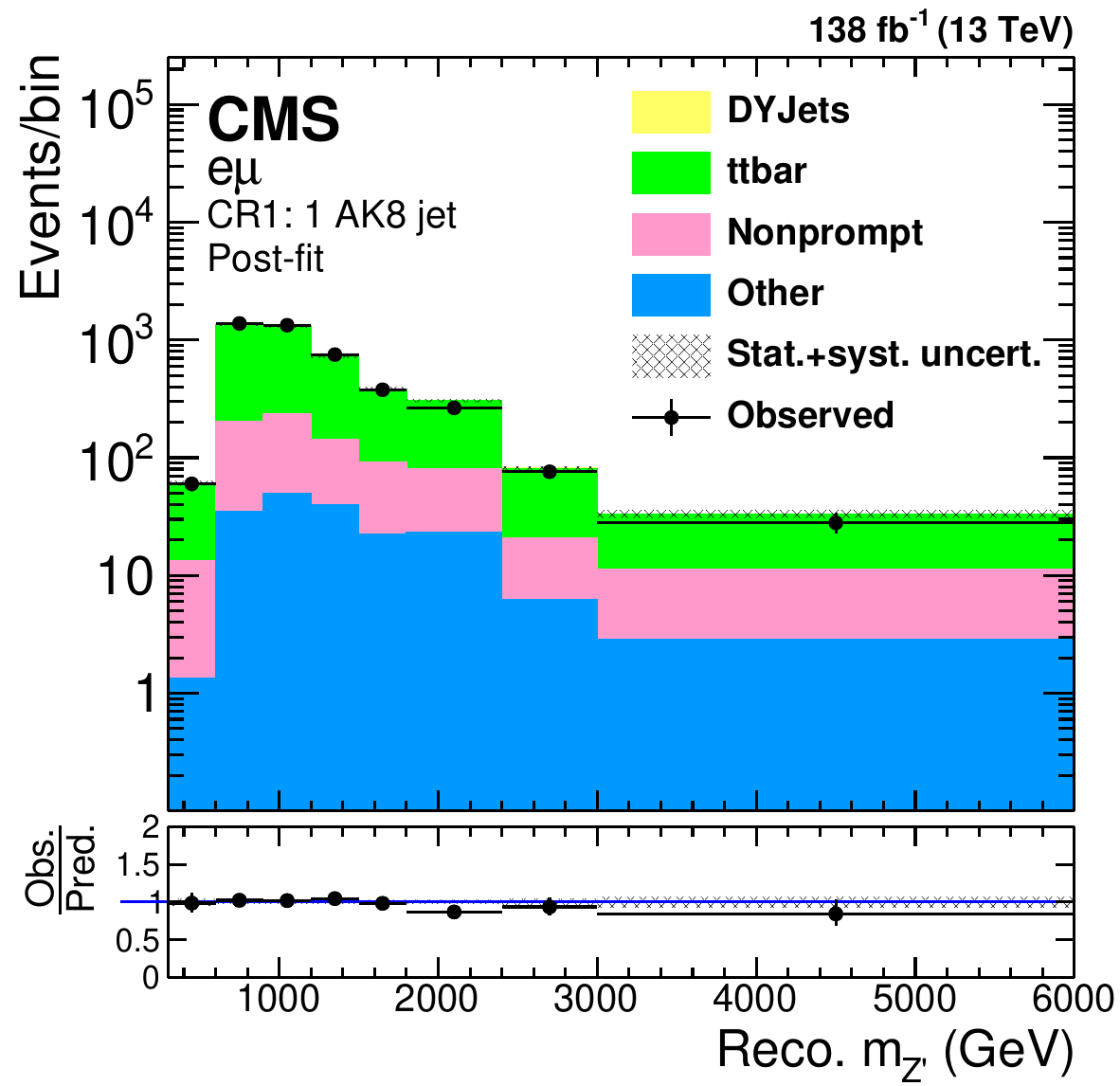}
    \hspace{0.01\textwidth}
    \includegraphics[width=0.45\textwidth]{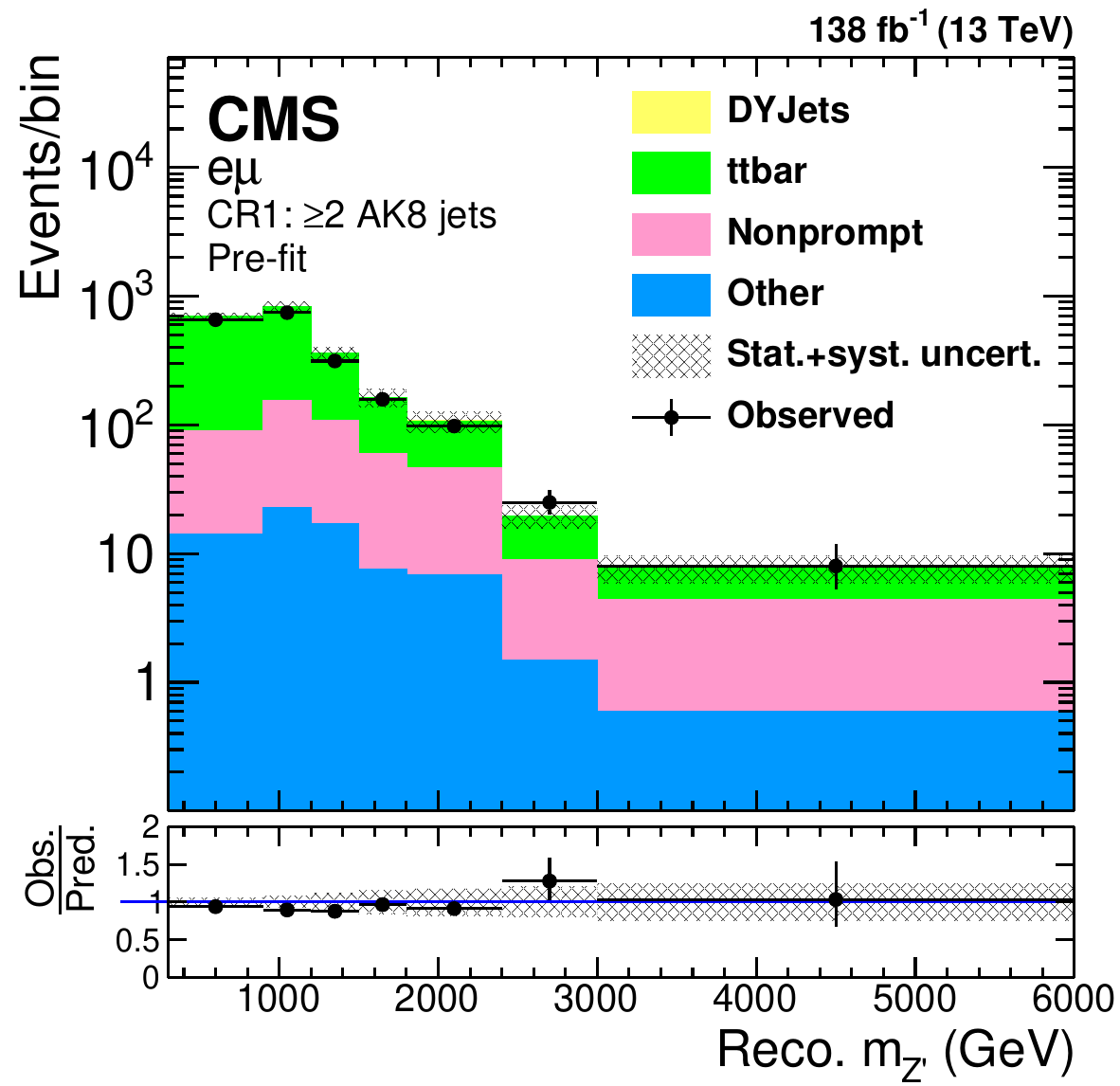}
    \hspace{0.01\textwidth}
    \includegraphics[width=0.45\textwidth]{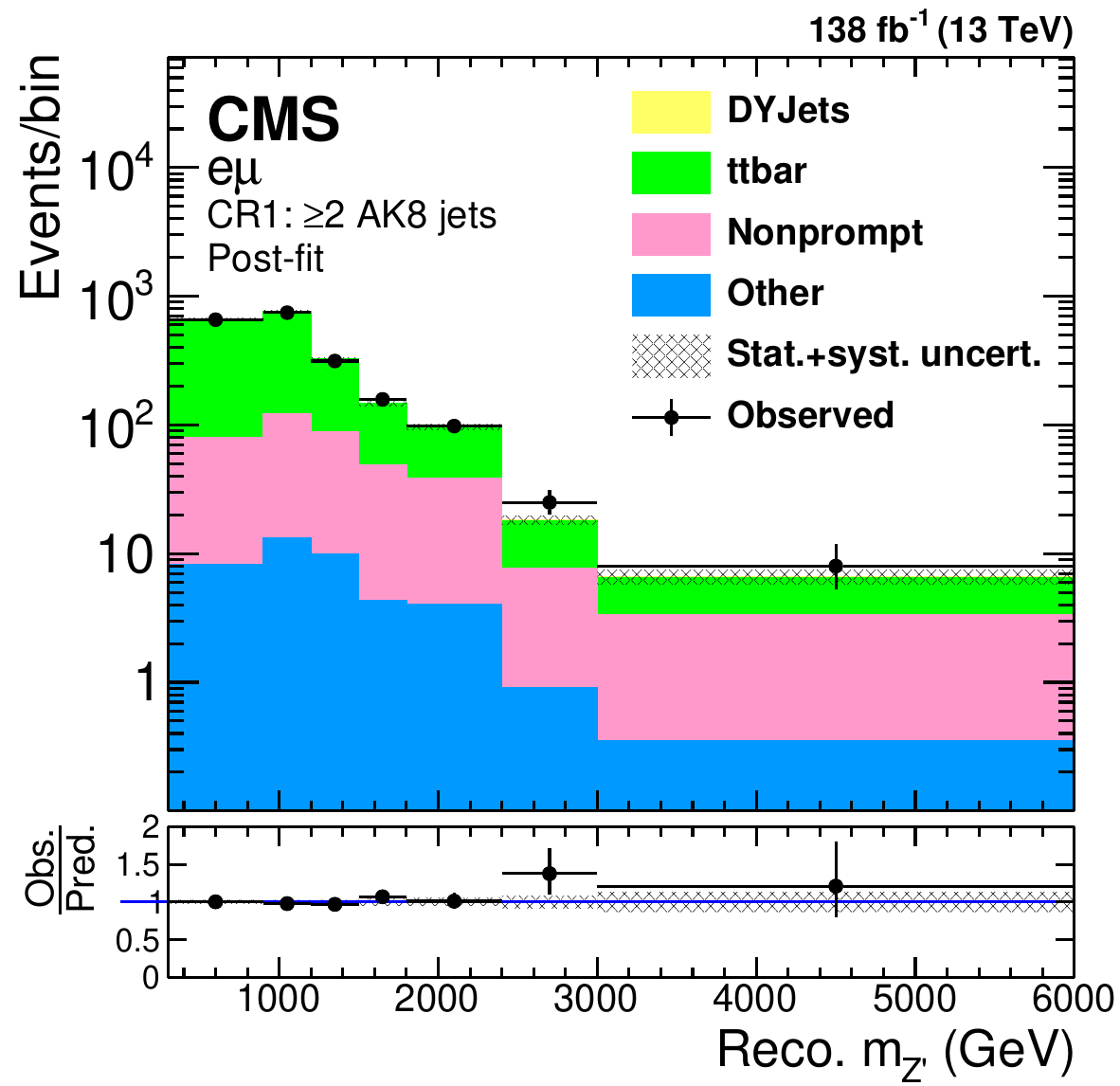}
    \caption{
        Reconstructed mass of the \PZpr candidate in CR1 ($\Pe\Pgm$), which consists of flavor sidebands of SR1 (upper), SR2 (middle), and SR3 (lower) regions.
        Pre-fit (post-fit) results are shown on the left (right). The fitting procedure is described in Section~\ref{sec:Systematics}.
        The last bin of each plot includes the overflow events.
        The contribution from DY events is too small to be visible on these plots.
        The lower panel of each plot shows the ratio of observed events to expected background.
}
    \label{fig:top_CRs}
\end{figure}

\begin{figure}[!hptb]
\centering
    \includegraphics[width=0.45\textwidth]{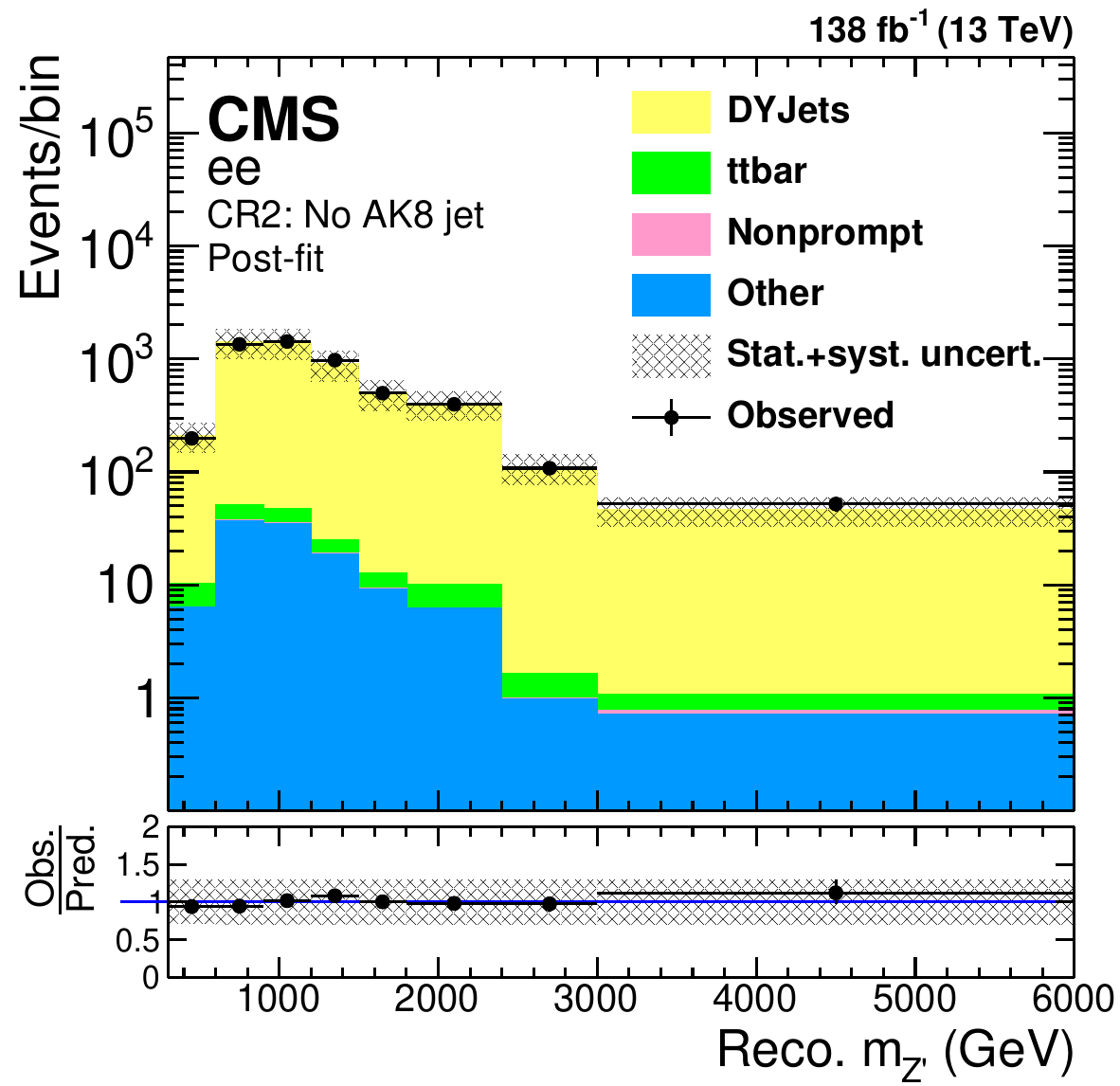}
    \hspace{0.01\textwidth}
    \includegraphics[width=0.45\textwidth]{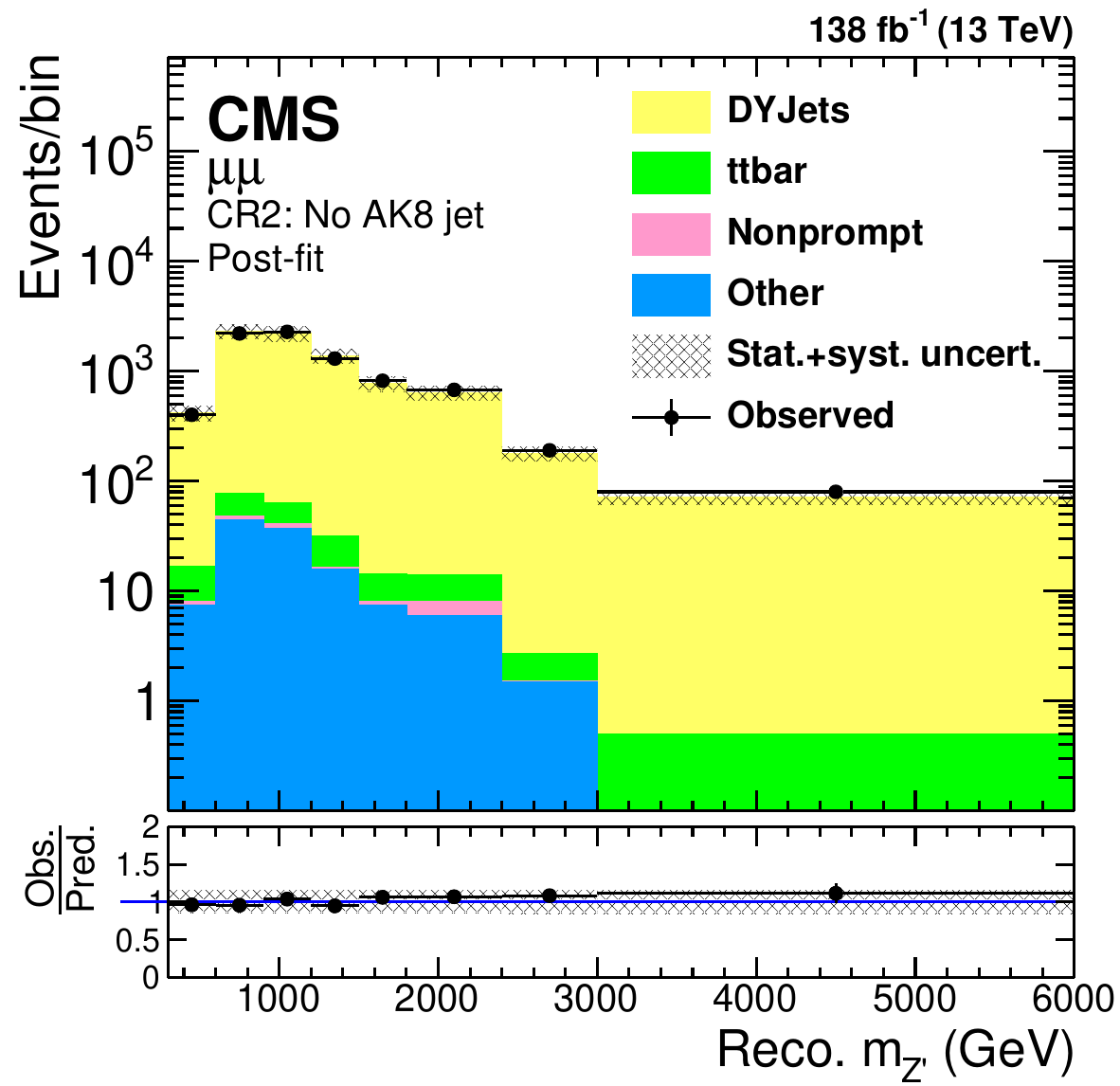}
    \hspace{0.01\textwidth}
    \includegraphics[width=0.45\textwidth]{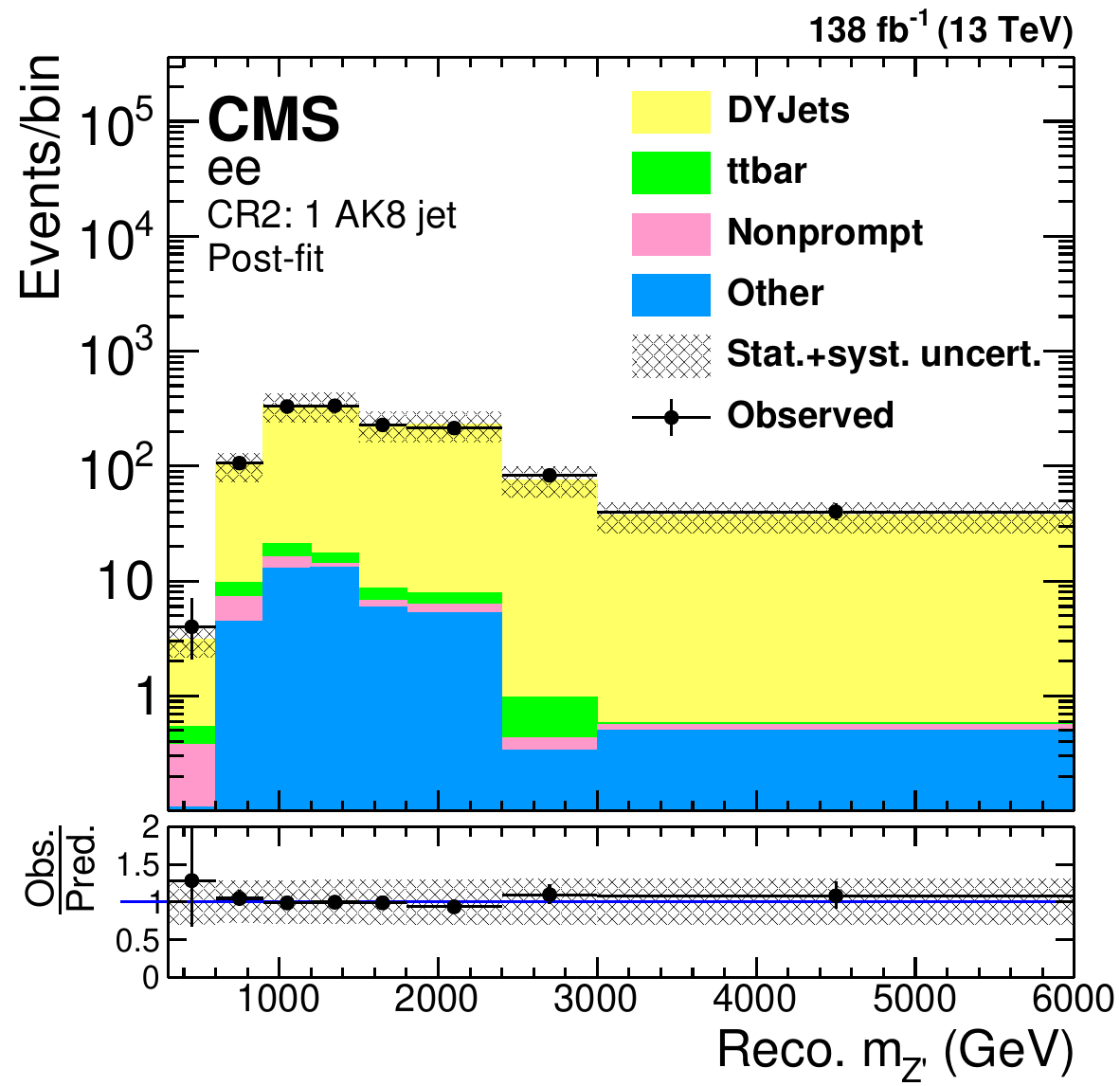}
    \hspace{0.01\textwidth}
    \includegraphics[width=0.45\textwidth]{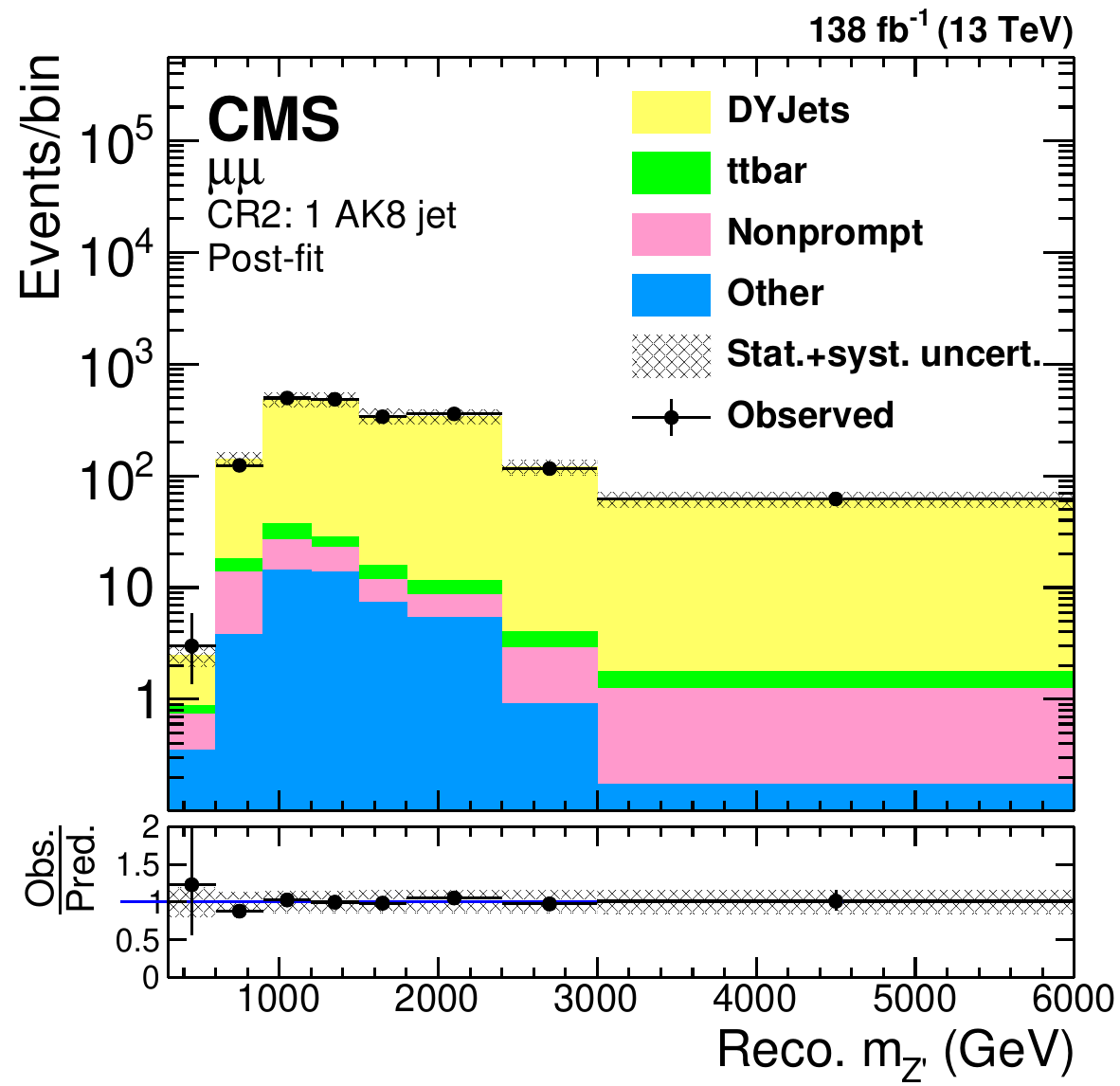}
    \hspace{0.01\textwidth}
    \includegraphics[width=0.45\textwidth]{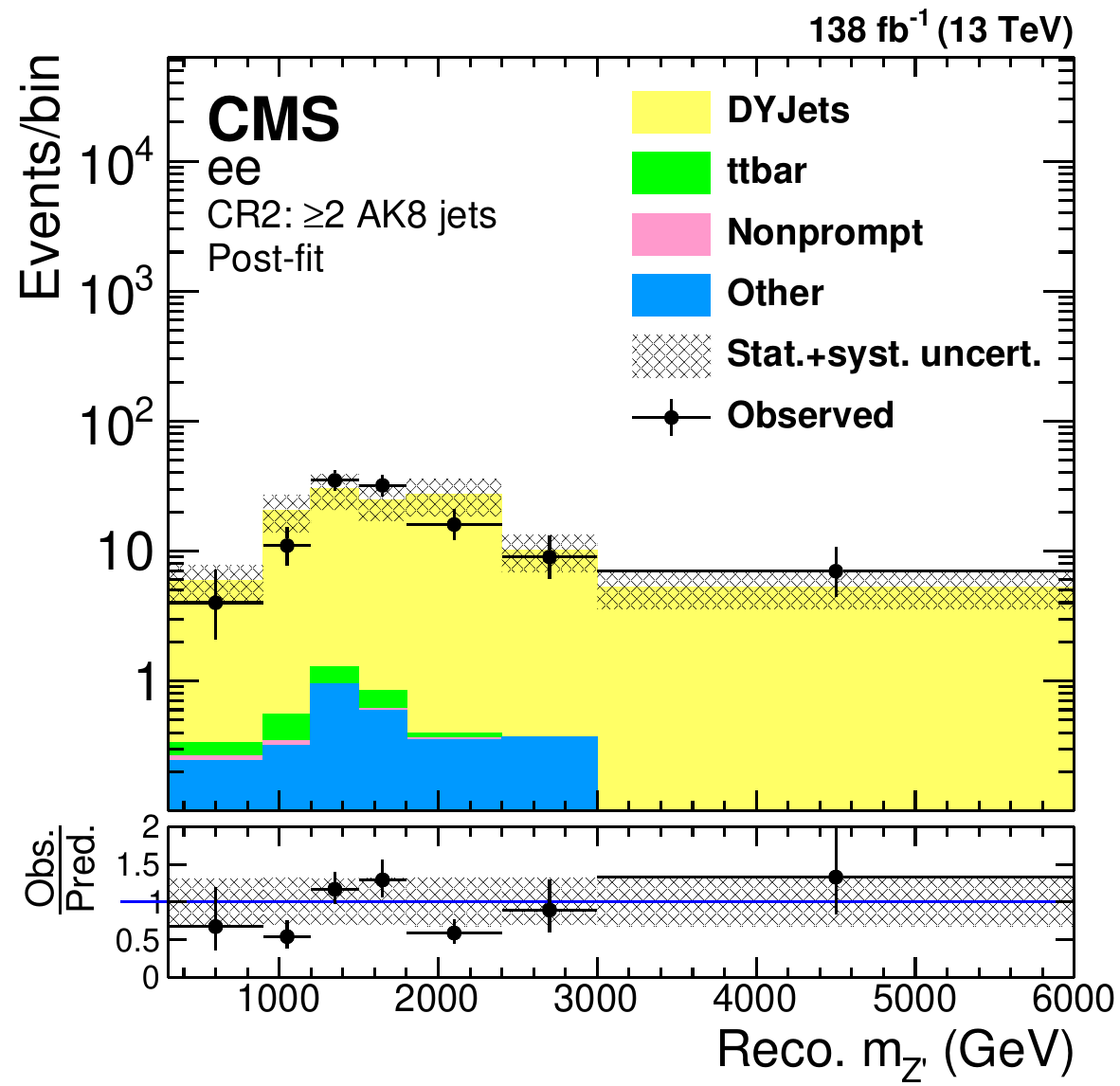}
    \hspace{0.01\textwidth}
    \includegraphics[width=0.45\textwidth]{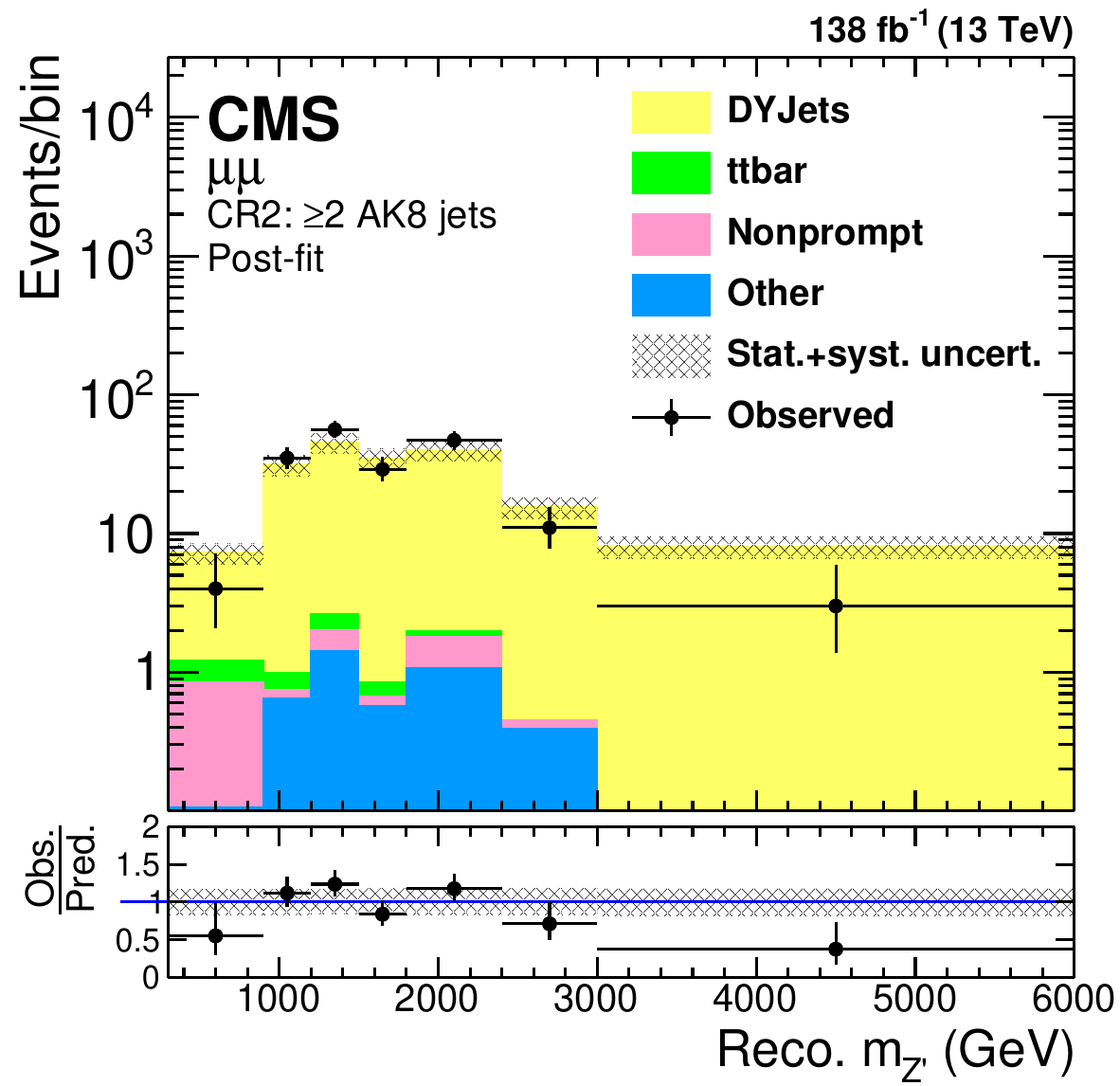}
    \caption{
        Reconstructed mass of the \PZpr candidate in CR2 ($\Pe\Pe$ and $\Pgm\Pgm$), which consists of the dilepton mass sidebands of SR1 (upper), SR2 (middle), and SR3 (lower) regions.
        Post-fit dielectron (dimuon) channel results are shown on the left (right). The fitting procedure is described in Section~\ref{sec:Systematics}.
        The last bin of each plot includes the overflow events.
        The lower panel of each plot shows the ratio of observed events to expected background.
}
    \label{fig:DY_CRs}
\end{figure}

Because of the large production cross section, DY events constitute the second largest background, although most of these are removed by the requirement, $m_{\ell\ell}>150\GeV$.
A second CR (CR2) is chosen to verify that the simulated sample of DY events is well modeled. The selection used to define CR2 is similar to that of the SR except the requirement on $m_{\ell\ell}$ is changed to $\abs{m_{\ell\ell} - {m}_{\PZ}} < 10\GeV$, where ${m}_{\PZ}$ is the mass of \PZ boson.
The events of CR2 are classified in three orthogonal regions according to the same criteria used for the SR, which are summarized in Table~\ref{tab:evt_sel_N_obj}.
To correct for the mismodeling of the dilepton \pt in simulated events,
a dedicated correction estimated using inclusive dimuon events is applied as a function of generator level dilepton \pt and mass~\cite{Sirunyan:2019bzr}.
The overall normalization factors as functions of the AK8 jet multiplicity are fit using all the CRs and SRs.
They range from 0.93 to 1.14 for events with fewer than two AK8 jets. For events with two or more AK8 jets, normalization factors have a wider range from 0.87 to 1.28 owing to the small numbers of events.
Figure~\ref{fig:DY_CRs} shows the agreement between the observed and simulated distributions of the reconstructed \PZpr candidate mass in CR2.

Additional prompt backgrounds are considered from di- and triboson production, \ttbar pair production in association with a vector boson, and single top production.
The contribution of these backgrounds is found to be almost negligible. We assign a conservative 50\% uncertainty in the estimated size of these backgrounds, to account for limited MC samples in the extreme phase space regions of interest.
The background contribution with nonprompt leptons, which are leptons not originating from the hard scattering process, is estimated using the simulated samples.
We assign an uncertainty of 100\% in the estimated size of the nonprompt-lepton background since this depends on detector effects, which are often more difficult to estimate.

\section{Systematic uncertainties and fitting procedure}
\label{sec:Systematics}
Binned maximum likelihood fits using the distribution of reconstructed \mZp for background and signal as templates are performed to extract the signal contribution for each SR.
The choice of bin size is based on the resolution for the reconstructed \mZp, which ranges from 150 to 550\GeV.
Where necessary, bins are then merged to ensure that the predicted number of MC events is nonzero in all bins of all SRs and CRs for each year.
Various sources of uncertainty in both shape and normalization of reconstructed \mZp distributions are considered as nuisance parameters and profiled in the fit.
Log-normal and Gaussian prior probability densities are used for normalization and shape uncertainties, respectively.
The normalizations of \ttbar and DY backgrounds are left freely floating in the fit.
The pre-fit and post-fit predictions on the plots of Figs.~\ref{fig:top_CRs}--\ref{fig:SR} correspond to the results obtained before and after likelihood fitting, respectively.
No nuisance parameter is pulled beyond one standard deviation from the pre-fit value.

For leptons, the efficiencies for reconstruction, identification, isolation, and trigger are estimated using observed and simulated $\PZ \to \Pe\Pe,\Pgm\Pgm$ events that are not included in the SRs.
Scale factors (SFs) are applied to simulated events to correct for mismodeling in the simulation.
Systematic uncertainties in momentum scale and resolution~\cite{CMS:2020uim,Sirunyan_2018} are estimated in simulations by varying the lepton momentum scales within their uncertainties.

The jet energy corrections and resolutions estimated using hadronic jets are used for both hadronic jets and AK8 jets with high \pt leptons inside.
Such corrections were found to provide a good modeling of jets also in the case of non-resolved lepton(s) within the AK8 jet in CRs.
The uncertainty in the jet energy scale and resolution are estimated using the same procedures as those used to determine systematic uncertainty in lepton momenta.
An additional uncertainty coming from the soft-drop mass scale is found to be less than 1.3\%~\cite{cms_jet_performance_8TeV}.

The integrated luminosities of the 2016, 2017, and 2018 data-taking periods are individually known with uncertainties in the range 1.2--2.5\%~\cite{CMS-LUM-17-003,CMS-PAS-LUM-17-004,CMS-PAS-LUM-18-002},
while the total 2016--2018 integrated luminosity has an uncertainty of 1.6\%, the improvement in precision reflecting the (uncorrelated) time evolution of some systematic effects.

The simulation of pileup events is based on a cross section of 69.2\unit{mb} for inelastic $\Pp\Pp$ collisions with energy deposits inside the CMS acceptance.
A systematic uncertainty for the pileup weight is assigned by varying the cross section up and down by 5\%.
An additional event weight that accounts for the L1 trigger inefficiency observed in the $\abs{\eta} < 2.0$ region for the 2016 and 2017 data-taking period is also applied~\cite{cms_level1_trigger_performance}.
The systematic uncertainty for the L1 trigger inefficiency weight is assigned by varying the weight within its uncertainty.

{\tolerance=800
The uncertainties related to the choice of PDF can affect both the cross section and kinematic distributions of the signal.
The effect of the PDF uncertainty and the variation related to the strong coupling constant (\alpS) are considered, following the method introduced by PDF4LHC15~\cite{Butterworth_2016}.
The uncertainties in the renormalization and factorization scales (${\mu}_{\mathrm{R}}$ and ${\mu}_{\mathrm{F}}$, respectively) are determined by varying them by a factor of two on an event-by-event basis
using combinations in which the two scales do not move in opposite directions at the same time.
The effect of these variations on the shape of the \mZp distribution is determined while maintaining the signal normalization.
\par}

The normalization uncertainties for the rare SM processes and nonprompt-lepton backgrounds are also incorporated into the likelihood as 
nuisance parameters.  We observe that the fit procedure returns an uncertainty of about 40\% in the nuisance parameter of nonprompt-lepton background, which is 
initially constrained at the 100\% level. Since this is consistent with the precision of independent background estimation methods using control samples in data-driven methods~\cite{CMS_HNL_dilep,PhysRevLett.120.221801,CMS:2016mku}, we use the fit to constrain the normalization.

The uncertainty in the dilepton \pt shape corrections to the simulated DY events is estimated by varying correction factors within their uncertainties.

A complete list of these uncertainties in the dielectron and dimuon channels is shown in Table~\ref{tab:syst_ee}.
The lepton trigger scale factors, jet energy resolution, L1 trigger inefficiency, and nonprompt lepton background normalizations are assumed to be uncorrelated between the 2016, 2017, and 2018 data sets,
since their corrections and corresponding uncertainties are affected by differences in operating conditions and detector performance between data-taking years.
Other experimental and all theoretical uncertainties are taken as correlated.

\begin{table}
\centering
\topcaption{Systematic uncertainties and their impacts on the total number of events in the signal regions before the binned maximum likelihood fit.}
\label{tab:syst_ee}
\cmsTable{
\begin{tabular}{l c c c c c c}
\multirow{2}{*}{Source} & \multirow{2}{*}{Bkgd./Signal process} & \multirow{2}{*}{Year-to-year treatment} & $\Pe\Pe$ bkgd. & $\Pe\Pe$ signal & $\Pgm\Pgm$ bkgd. & $\Pgm\Pgm$ signal \\
&                                       &     &  (\%)           & (\%)            & (\%)           & (\%) \\
\hline
Integrated luminosity & All bkgd./Signal & Correlated & 1.6 & 1.6 & 1.6 & 1.6\\ 
Electron momentum scale & All bkgd./Signal & Correlated & 0.1--0.9 & 0.0--0.5 &  $<$0.1  &  $<$0.1 \\
Electron momentum smear & All bkgd./Signal & Correlated &  $<$0.1  & 0.0--0.1 &  $<$0.1  &  $<$0.1 \\
Muon momentum scale & All bkgd./Signal & Correlated &  $<$0.1  & 0.0--0.1 & 0.0--5.0 & 0.0--1.1\\
Jet energy resolution & All bkgd./Signal & Uncorrelated & 1.4-- 14.4 & 0.0--10.8 & 0.5--14.8 & 0.0--9.5\\
Jet energy scale & All bkgd./Signal & Correlated & 5.4--15.3 & 0.0--11.4 & 3.6--16.7 & 0.0--9.2\\
Soft-drop mass scale & All bkgd./Signal & Correlated & 0.0--0.5 & 0.0--1.3 & 0.0--0.5 & 0.0--1.2\\
Pileup weight & All bkgd./Signal & Correlated & 0.0--2.0 & 0.0--2.9 & 0.0--1.7 & 0.0--1.9\\
L1 trigger inefficiency & All bkgd./Signal & Uncorrelated & 0.0--1.8 & 0.0--2.3 & 0.0--1.1 & 0.0--1.0\\
Electron reconstruction SF & All bkgd./Signal & Correlated & 1.3--2.2 & 1.1--2.3 &  $<$0.1  &  $<$0.1 \\
Electron ident. SF & All bkgd./Signal & Correlated & 1.9--4.0 & 0.3--5.3 &  $<$0.1  &  $<$0.1 \\
Electron trigger SF & All bkgd./Signal & Uncorrelated & 0.4--2.8 & 0.1--15.1 &  $<$0.1  &  $<$0.1 \\
Muon reconstruction SF & All bkgd./Signal & Correlated &  $<$0.1  &  $<$0.1  & 0.3--1.0 & 0.3--10.4\\
Muon ident. SF & All bkgd./Signal & Correlated &  $<$0.1  &  $<$0.1  & 0.2--1.6 & 0.2--1.6\\
Muon isolation SF & All bkgd./Signal & Correlated &  $<$0.1  &  $<$0.1  &  $<$0.1  &  $<$0.1 \\
Muon trigger SF & All bkgd./Signal & Uncorrelated &  $<$0.1  &  $<$0.1  & 0.1--0.2 & 0.1--1.3\\
Dilepton \pt shape correction & DY$+$jets & Correlated & 0.0--19.2 & \NA & 0.0--18.8 & \NA\\
PDF & Signal & Correlated & \NA &  $<$0.1 & \NA & $<$0.1 \\
Scale (${\mu}_{\mathrm{R}}, {\mu}_{\mathrm{F}}$)  & Signal & Correlated & \NA &  $<$0.1 & \NA & $<$0.1 \\
Nonprompt norm. & Nonprompt & Uncorrelated & 100 & \NA & 100 & \NA  \\
Rare SM norm. & Others & Correlated & 50 & \NA & 50 & \NA \\
\end{tabular}
}
\end{table}

\section{Results and discussion}
\label{sec:Results}
\begin{figure}[!hptb]
\centering
    \includegraphics[width=0.45\textwidth]{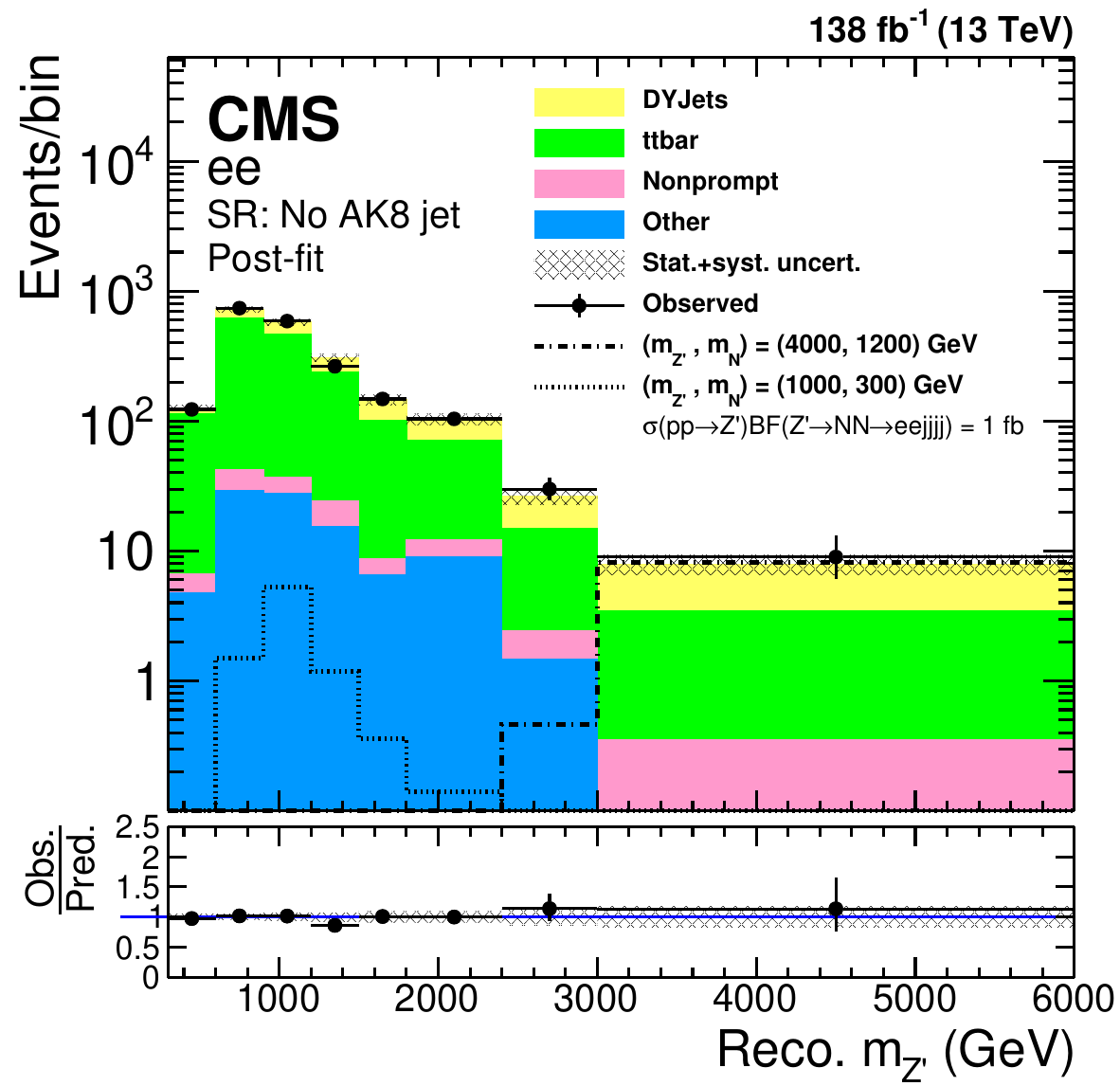}
    \hspace{0.01\textwidth}
    \includegraphics[width=0.45\textwidth]{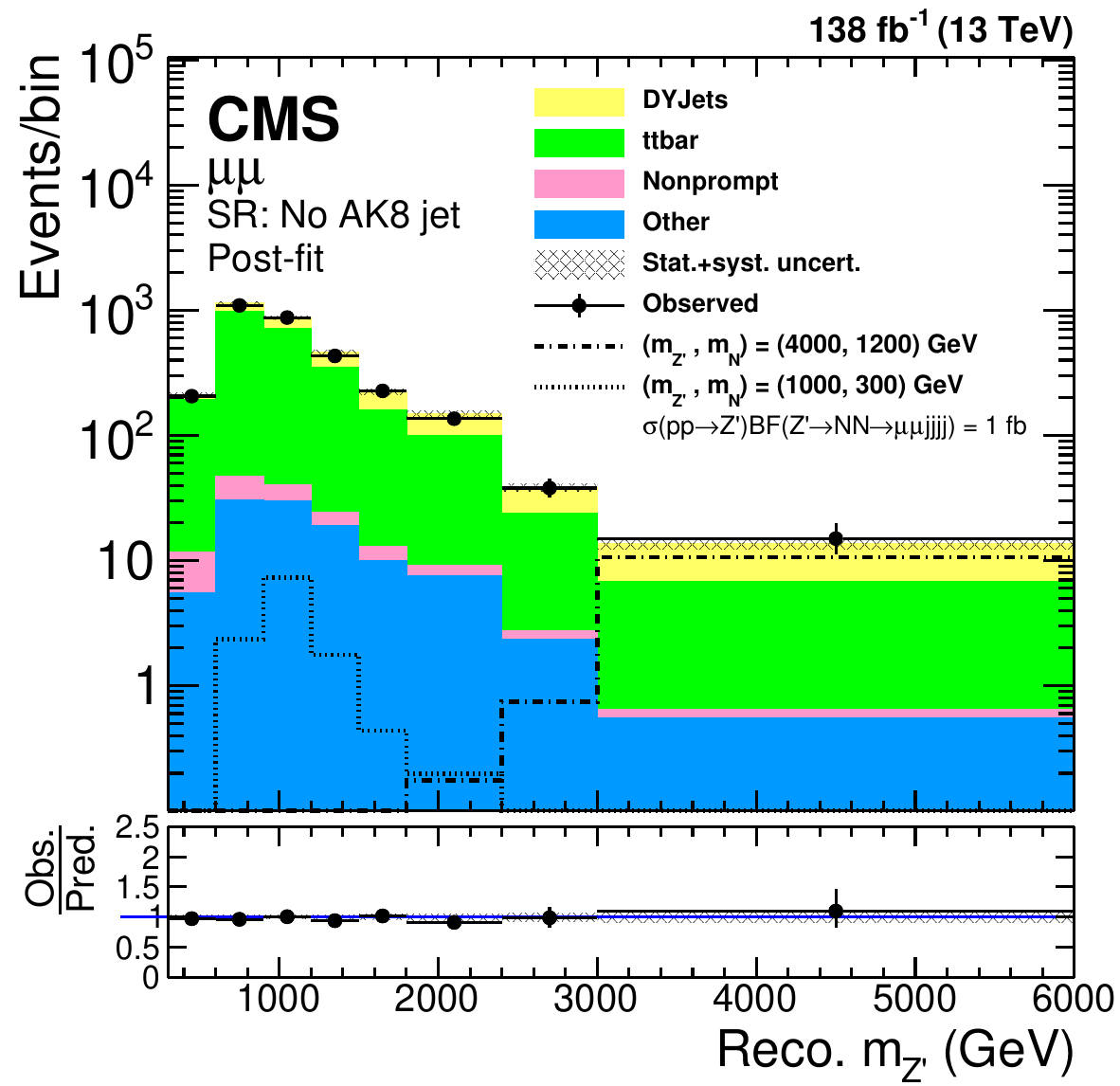}
    \hspace{0.01\textwidth}
    \includegraphics[width=0.45\textwidth]{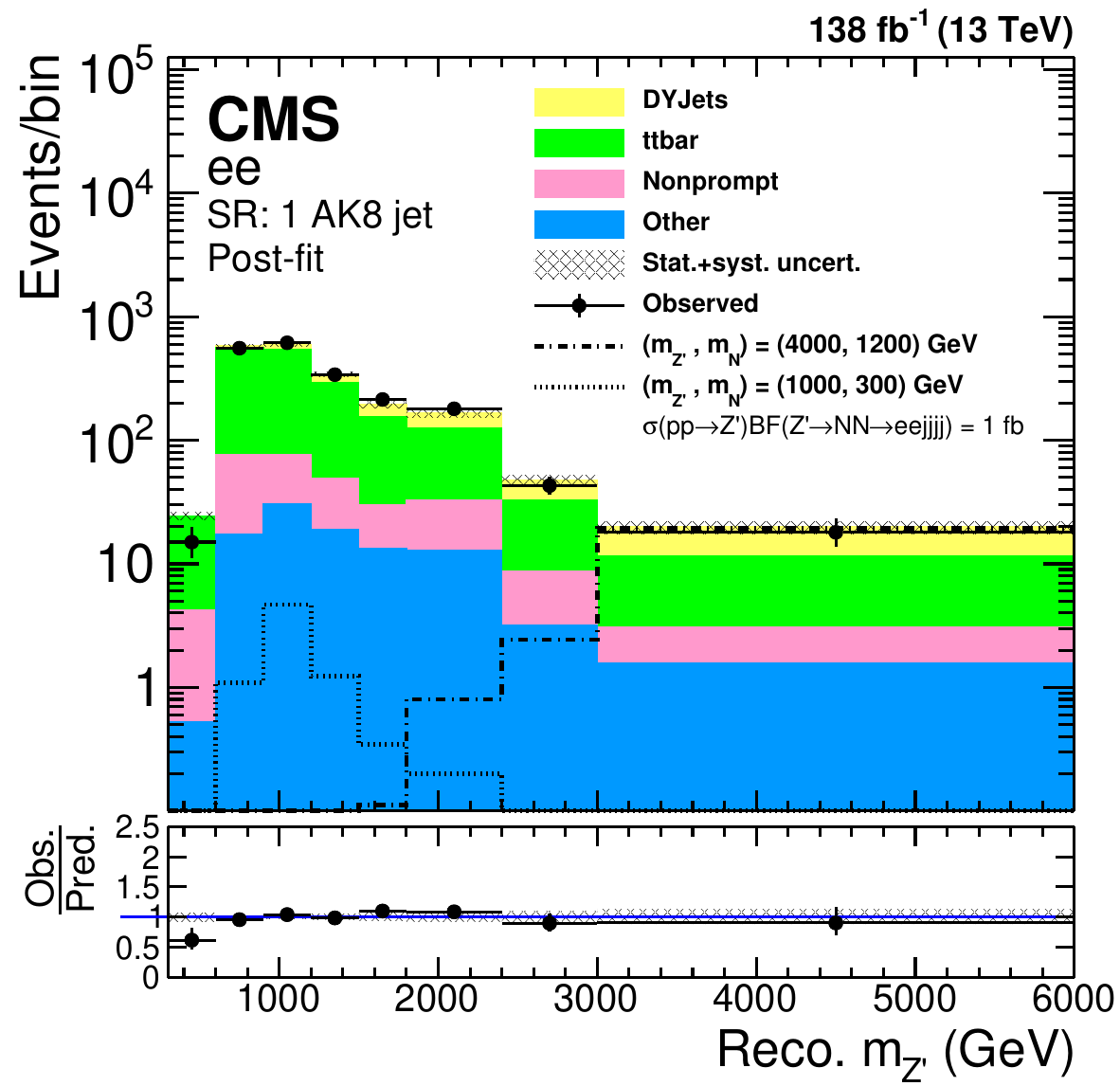}
    \hspace{0.01\textwidth}
    \includegraphics[width=0.45\textwidth]{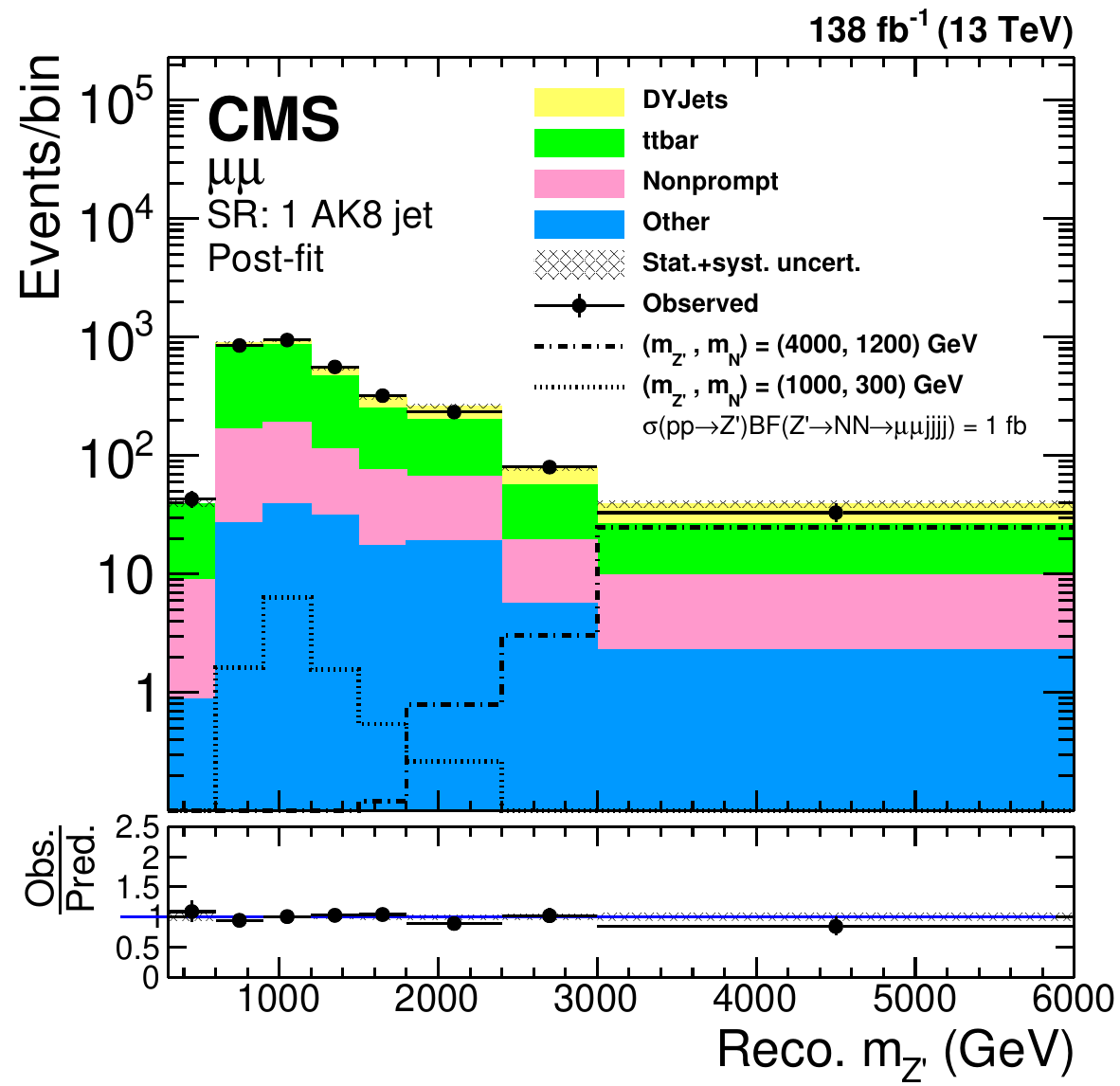}
    \hspace{0.01\textwidth}
    \includegraphics[width=0.45\textwidth]{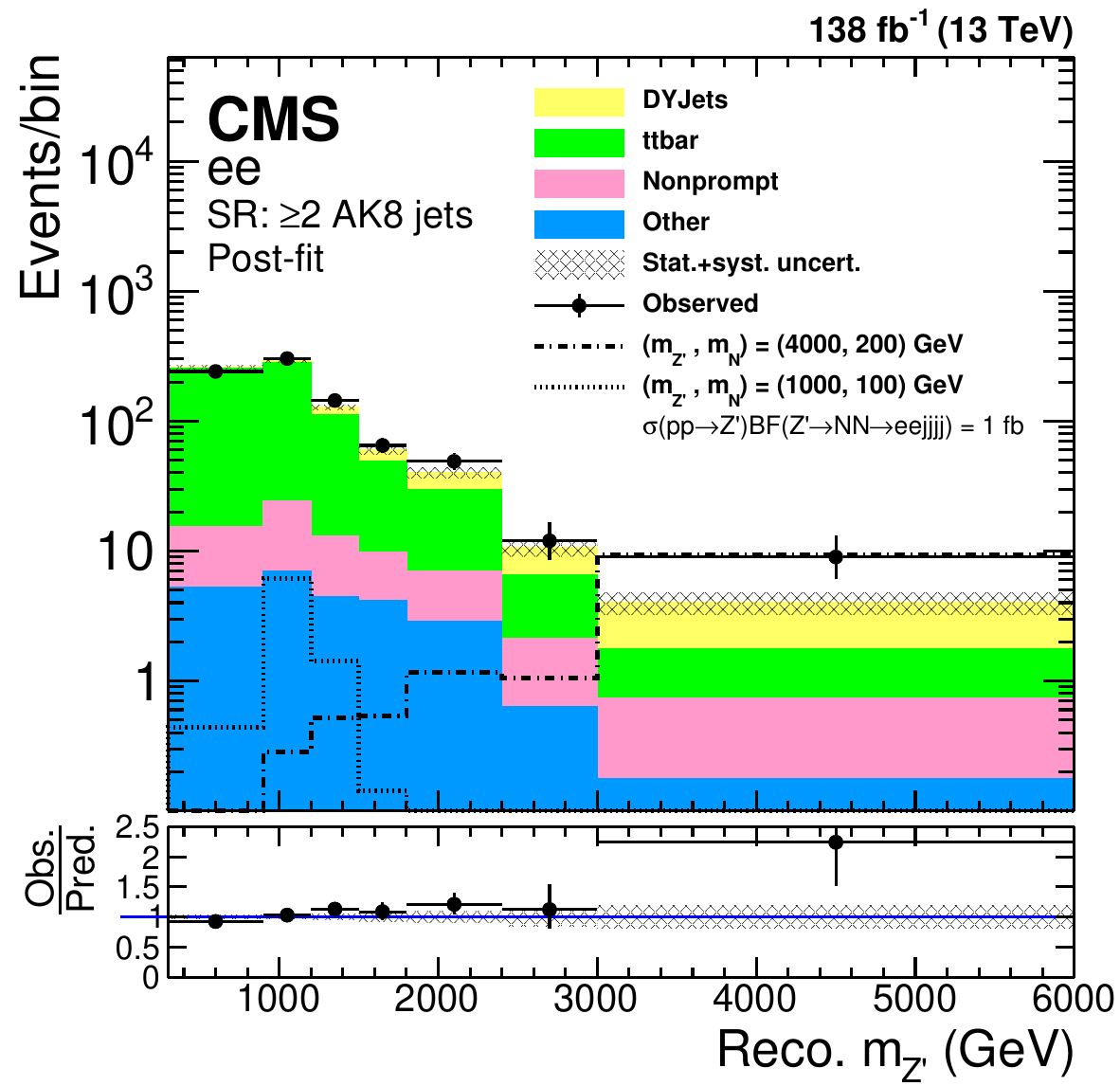}
    \hspace{0.01\textwidth}
    \includegraphics[width=0.45\textwidth]{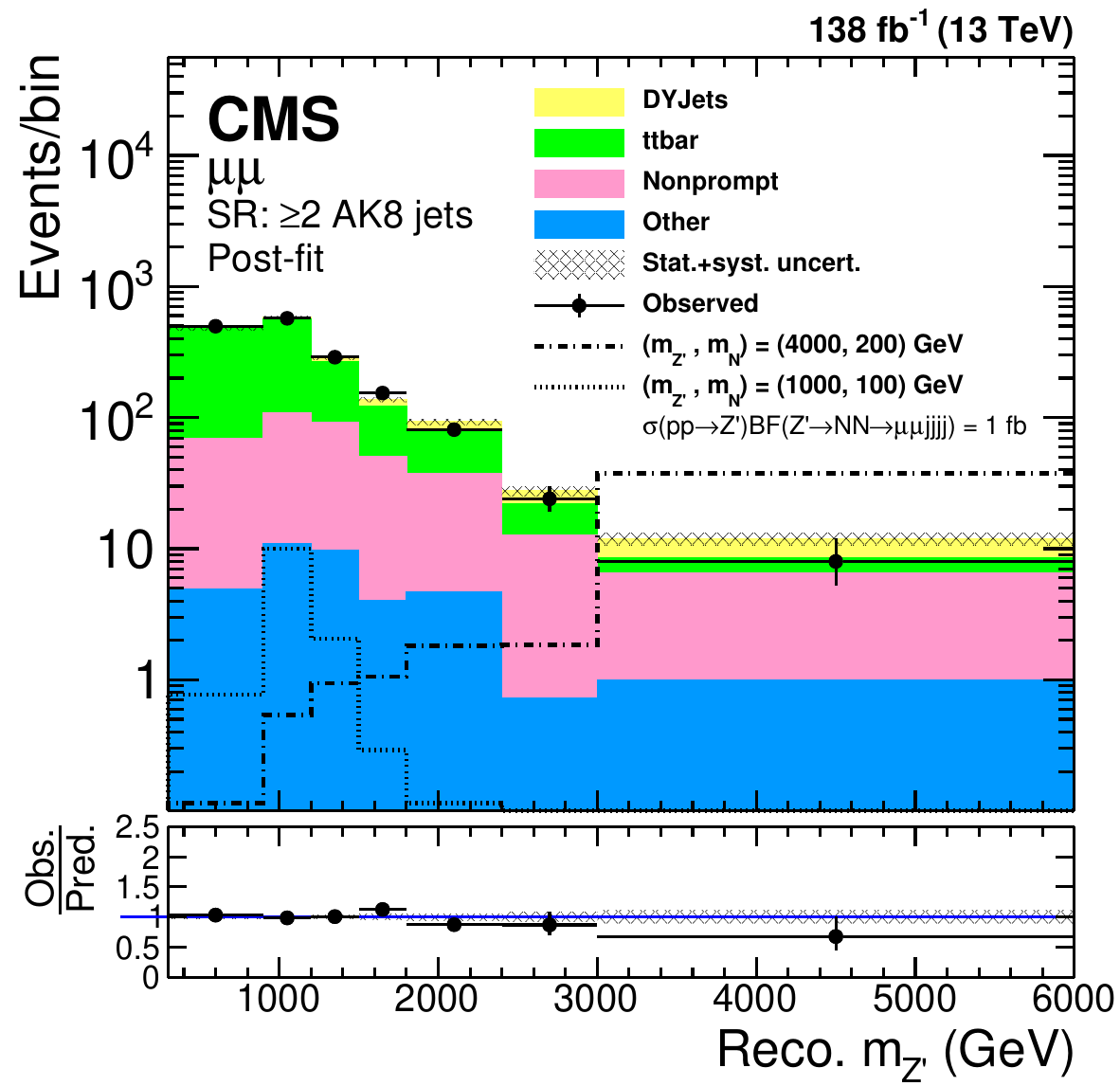}
    \caption{
    Distributions of the reconstructed \PZpr candidate mass in SR1 (upper row), SR2 (middle row) and SR3 (lower row). The left (right) column corresponds to the dielectron (dimuon) channel.
    Signal samples with $(\mZp, \mN) = (1000, 100)\GeV$, $(1000, 300)\GeV$, $(4000, 200)\GeV$, and $(4000, 1200)\GeV$
    are shown together with a reference cross section times branching fraction of 1\unit{fb}.
    The last bin of each plot includes the overflow events.
    The lower panel of each plot shows the ratio of observed events to expected background.
}
    \label{fig:SR}
\end{figure}

The background-only post-fit invariant mass distributions of \PZpr candidates in the different SRs are shown in Fig.~\ref{fig:SR}.
Representative signal distributions are also shown.

We do not observe any significant excess in data with respect to the SM prediction.
The 95\% confidence level (\CL) upper limits on the product of signal cross section and BF are obtained
using the distributions of the likelihood ratio calculated with the asymptotic approximation~\cite{Cowan:2010js} and the \CLs criterion~\cite{Junk:1999kv, Read:2002hq}.

\begin{figure}[!hptb]
\centering
    \includegraphics[width=0.48\textwidth]{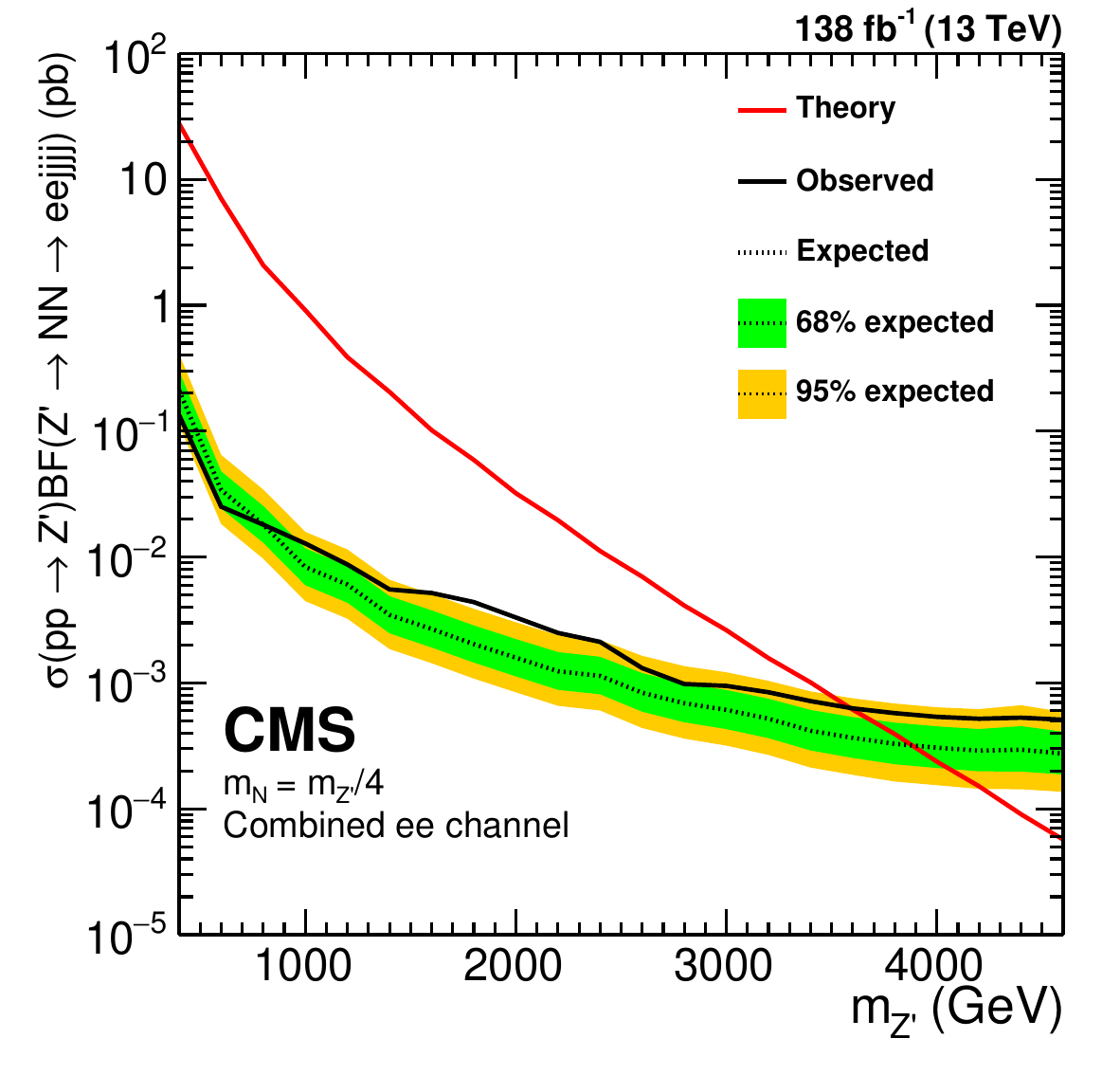}
    \hspace{0.01\textwidth}
    \includegraphics[width=0.48\textwidth]{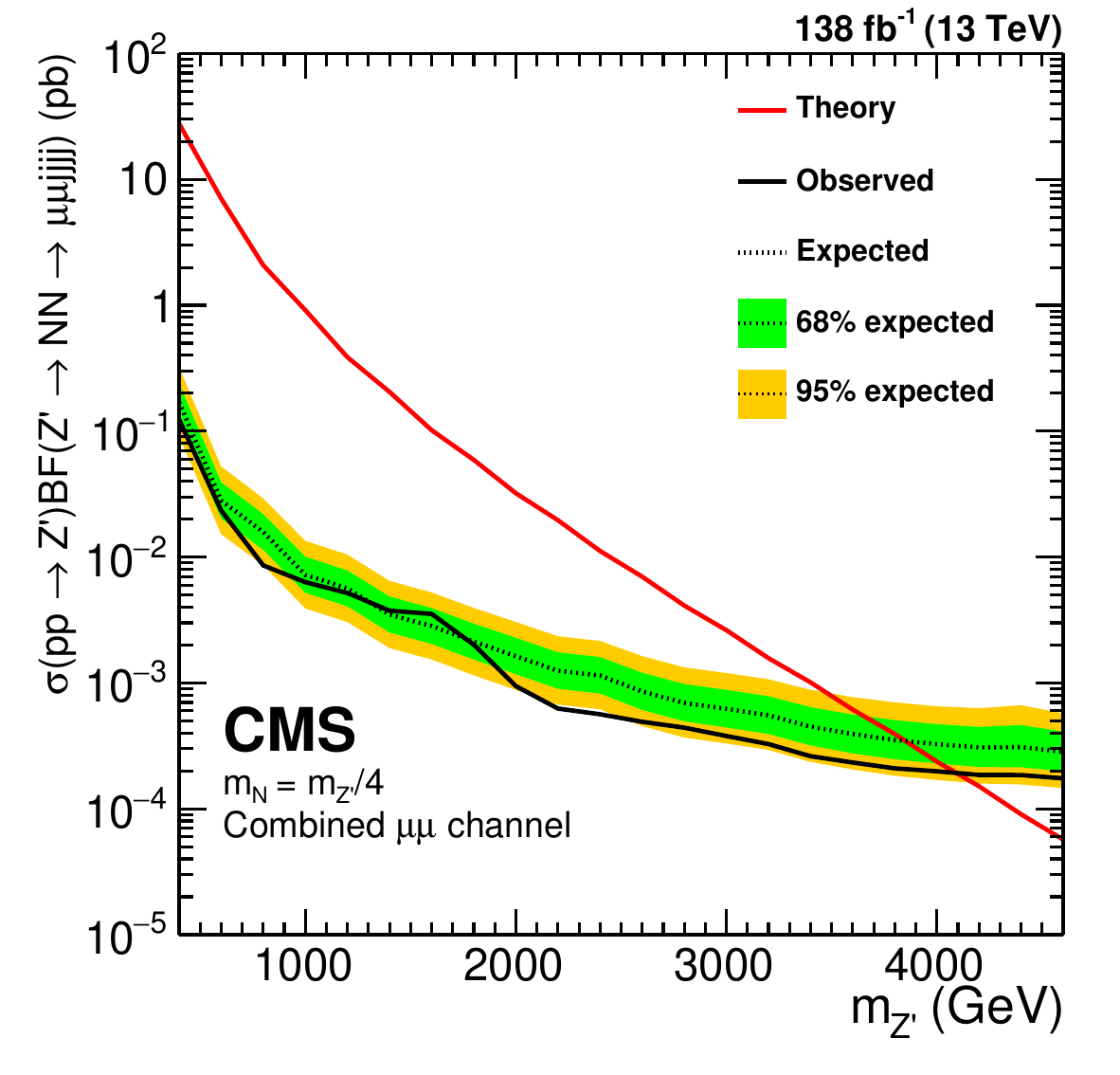}
    \hspace{0.01\textwidth}
    \includegraphics[width=0.48\textwidth]{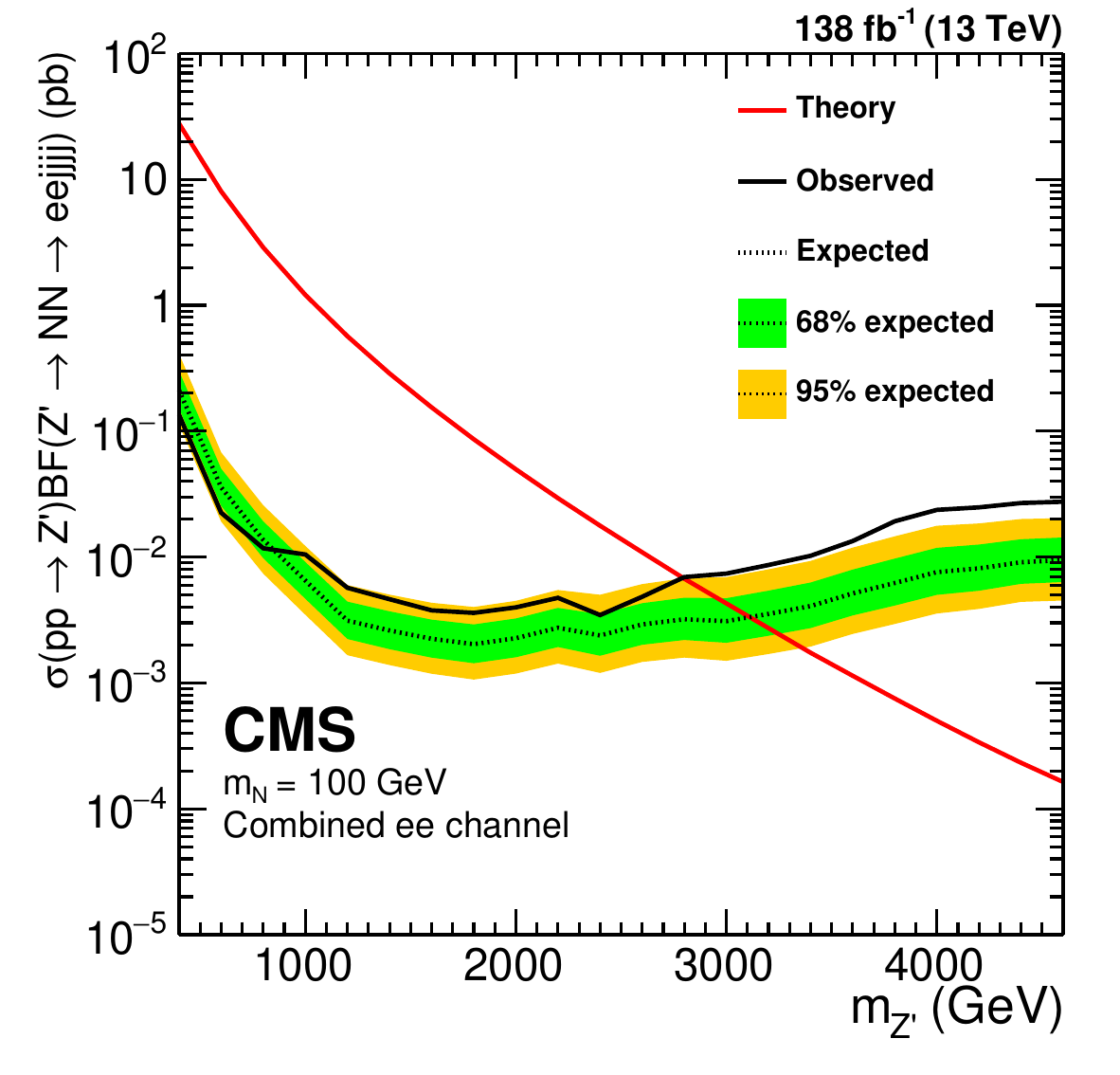}
    \hspace{0.01\textwidth}
    \includegraphics[width=0.48\textwidth]{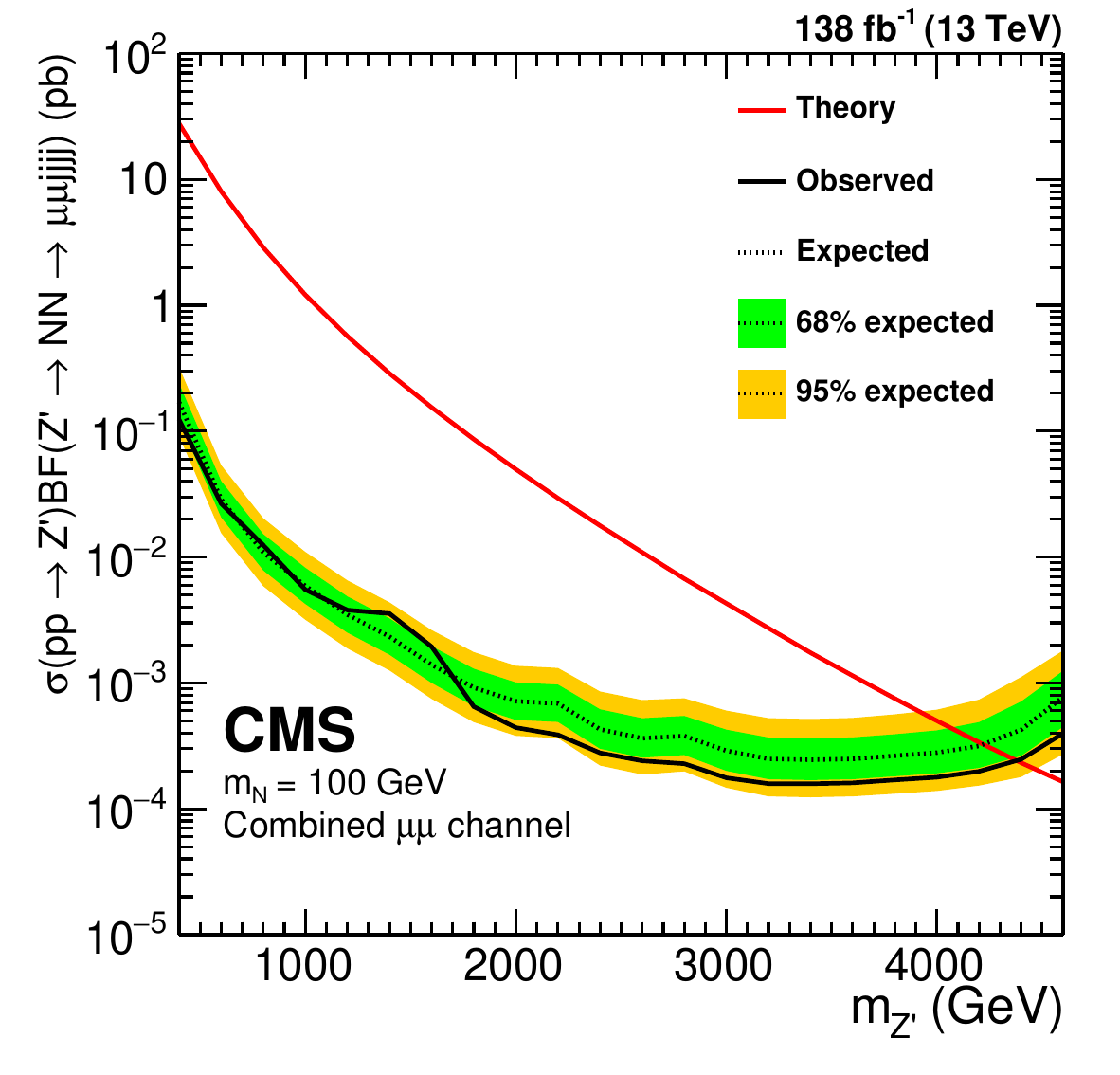}
    \caption{
    The observed and expected 95\% \CL upper limits on the product of \PZpr boson cross section and branching fraction are shown
    for the case of $\mN = \mZp / 4$ (upper row) and $\mN = 100\GeV$ (lower row) for dielectron (left column) and dimuon (right column) channels.
    The green and yellow bands indicate the 68\% and 95\% \CL regions around the expected limit.
    The red lines represent values coming from the benchmark LRSM model~\cite{Mattelaer:2016ynf}.
    Both opposite- and same-sign dileptons from decays of heavy neutrino pairs are considered.
    }
    \label{fig:1D_limits}
\end{figure}

\begin{figure}[!hptb]
\centering
    \includegraphics[width=0.48\linewidth]{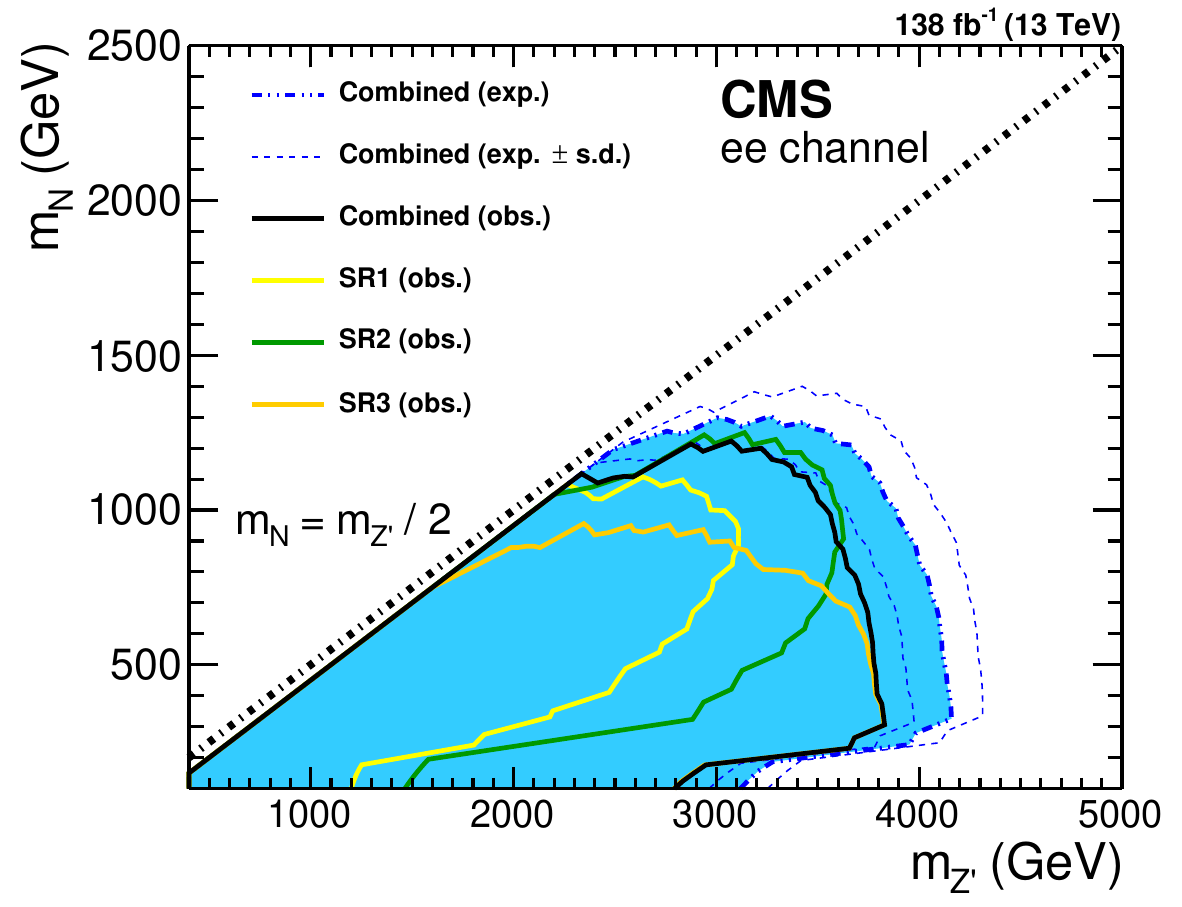}
    \hspace{0.01\textwidth}
    \includegraphics[width=0.48\linewidth]{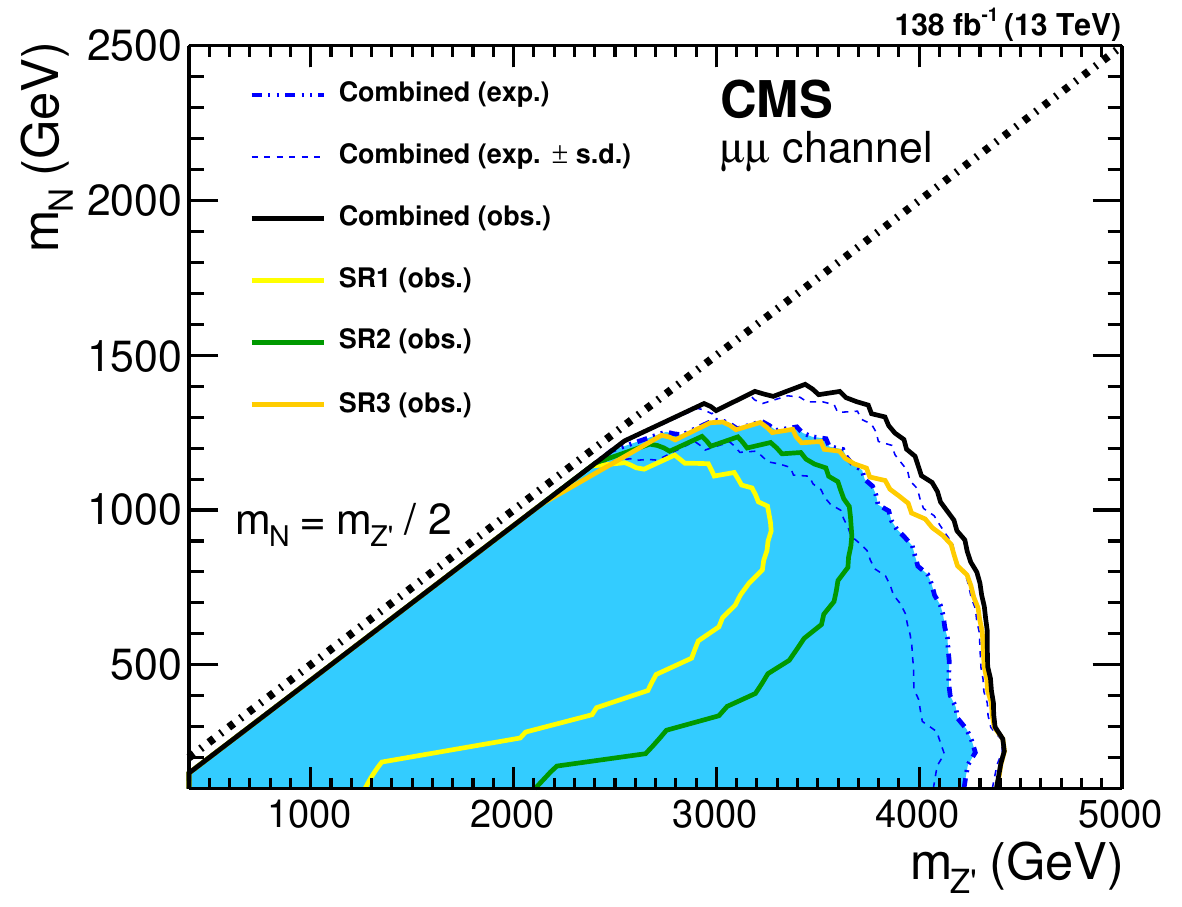}
    \caption{
    Observed and expected exclusion regions at 95\% \CL in the 2D phase space of \mZp vs. \mN for dielectron (left) and dimuon (right) channels. One standard deviation (s.d.) limits are also shown.
    Both opposite- and same-sign dileptons from decays of heavy neutrino pairs are considered.
    }   
    \label{fig:limit_ElElMuMu}
\end{figure}

The observed and expected exclusion limits on the signal cross section times BF are shown in Fig.~\ref{fig:1D_limits}.
The maximum local significance is $3.32\sigma$, observed in the dielectron channel for the $(\mZp, \mN) = (4.6, 0.1)\TeV$.
The probability to observe similar or larger excess in the dielectron channel across the full analysis mass range is estimated with pseudo experiments under background-only hypothesis. This probability is $1.12 \times {10}^{-2}$, corresponding to a global significance of $2.28\sigma$.
For cases where $\mN = \mZp / 4$, the observed (expected) lower limits at 95\% \CL on the mass of the \PZpr boson are 3.59 (3.90)\TeV in the dielectron channel and 4.10 (3.86)\TeV in the dimuon channel.
For the phase space with mostly boosted signals ($\mN = 100\GeV$),
the observed (expected) lower limits are found to be $\mZp = 2.79\,(3.12)\TeV$ and 4.38 (4.22)\TeV in the dielectron and dimuon channels, respectively.
The use of a dedicated algorithm for boosted signals provides a significant improvement in sensitivity, in particular for the muon channel.
We note that the sensitivity for Dirac-type heavy neutrinos is identical as long as the branching fraction of the \PZpr boson to a heavy neutrino pair stays the same.

One of the assumptions in this analysis is the presence of a nonnegligible \PZpr coupling to quarks.
A comparison with the sensitivity of the search for high mass dijet resonances performed by CMS using the same data set~\cite{CMS:2019gwf} is therefore possible.
For comparable \PZpr branching fractions into quarks and heavy neutrinos, our search is found to be about one order of magnitude more sensitive.

The upper limits across the $\mN$-$\mZp$ plane are shown in Fig.~\ref{fig:limit_ElElMuMu}.
The SR1 is sensitive in the phase space regions $\mZp = 1$ to 2\TeV and $\mN = 0.5$ to 1\TeV. The SR2 is particularly sensitive for higher \mZp with resolved heavy neutrinos.
The SR3 is most important where heavy neutrinos are boosted.
The dimuon channel shows better sensitivity where heavy neutrinos are mostly boosted because the dielectron channel suffers for the H/E requirements discussed in Section~\ref{sec:EvtRecoID}.
In the dielectron channel, for $\mN > 0.9\TeV$ and $\mZp > 2.85\TeV$, the observed limits obtained in the SR2 are better than those of the combined limits.
This indicates that the slight excess observed for the combined limits in this region of phase space is driven by the SR3.

A priori, a further categorization separating same-sign and opposite-sign dilepton pair events might increase the sensitivity.
However, with the current data sample size, the evaluation of same-sign dilepton backgrounds using control samples, particularly in a highly boosted regime for heavy neutrinos,
is affected by large systematic uncertainties. Thus introducing sign categorization would not improve the present search.

\section{Summary}
\label{sec:Summary}
A search for the pair production of heavy Majorana neutrinos (\N) via the decay of a \PZpr boson, in a final state with two same-flavor leptons and at least two reconstructed jets has been performed in proton-proton collisions at a center-of-mass energy of 13\TeV, using LHC 2016--2018 data corresponding to an integrated luminosity of 138\fbinv.
No significant excess of events beyond the expected background is observed.
Upper limits on the product of signal cross section and branching fraction in the context of a left-right symmetry model scenario are set~\cite{Mattelaer:2016ynf}.
Exclusion regions in the dielectron and dimuon channels are set at 95\% confidence level.
These are the first results of a search for this process at 13\TeV and are the most restrictive limits to date on the mass of \N as a function of the \PZpr boson mass.

\begin{acknowledgments}
We congratulate our colleagues in the CERN accelerator departments for the excellent performance of the LHC and thank the technical and administrative staffs at CERN and at other CMS institutes for their contributions to the success of the CMS effort. In addition, we gratefully acknowledge the computing centers and personnel of the Worldwide LHC Computing Grid and other centers for delivering so effectively the computing infrastructure essential to our analyses. Finally, we acknowledge the enduring support for the construction and operation of the LHC, the CMS detector, and the supporting computing infrastructure provided by the following funding agencies: SC (Armenia), BMBWF and FWF (Austria); FNRS and FWO (Belgium); CNPq, CAPES, FAPERJ, FAPERGS, and FAPESP (Brazil); MES and BNSF (Bulgaria); CERN; CAS, MoST, and NSFC (China); MINCIENCIAS (Colombia); MSES and CSF (Croatia); RIF (Cyprus); SENESCYT (Ecuador); MoER, ERC PUT and ERDF (Estonia); Academy of Finland, MEC, and HIP (Finland); CEA and CNRS/IN2P3 (France); BMBF, DFG, and HGF (Germany); GSRI (Greece); NKFIH (Hungary); DAE and DST (India); IPM (Iran); SFI (Ireland); INFN (Italy); MSIP and NRF (Republic of Korea); MES (Latvia); LAS (Lithuania); MOE and UM (Malaysia); BUAP, CINVESTAV, CONACYT, LNS, SEP, and UASLP-FAI (Mexico); MOS (Montenegro); MBIE (New Zealand); PAEC (Pakistan); MES and NSC (Poland); FCT (Portugal); MESTD (Serbia); MCIN/AEI and PCTI (Spain); MOSTR (Sri Lanka); Swiss Funding Agencies (Switzerland); MST (Taipei); MHESI and NSTDA (Thailand); TUBITAK and TENMAK (Turkey); NASU (Ukraine); STFC (United Kingdom); DOE and NSF (USA).
   
\hyphenation{Rachada-pisek} Individuals have received support from the Marie-Curie program and the European Research Council and Horizon 2020 Grant, contract Nos.\ 675440, 724704, 752730, 758316, 765710, 824093, 884104, and COST Action CA16108 (European Union); the Leventis Foundation; the Alfred P.\ Sloan Foundation; the Alexander von Humboldt Foundation; the Science Committee, project no. 22rl-037 (Armenia); the Belgian Federal Science Policy Office; the Fonds pour la Formation \`a la Recherche dans l'Industrie et dans l'Agriculture (FRIA-Belgium); the Agentschap voor Innovatie door Wetenschap en Technologie (IWT-Belgium); the F.R.S.-FNRS and FWO (Belgium) under the ``Excellence of Science -- EOS" -- be.h project n.\ 30820817; the Beijing Municipal Science \& Technology Commission, No. Z191100007219010; the Ministry of Education, Youth and Sports (MEYS) of the Czech Republic; the Hellenic Foundation for Research and Innovation (HFRI), Project Number 2288 (Greece); the Deutsche Forschungsgemeinschaft (DFG), under Germany's Excellence Strategy -- EXC 2121 ``Quantum Universe" -- 390833306, and under project number 400140256 - GRK2497; the Hungarian Academy of Sciences, the New National Excellence Program - \'UNKP, the NKFIH research grants K 124845, K 124850, K 128713, K 128786, K 129058, K 131991, K 133046, K 138136, K 143460, K 143477, 2020-2.2.1-ED-2021-00181, and TKP2021-NKTA-64 (Hungary); the Council of Science and Industrial Research, India; the Latvian Council of Science; the Ministry of Education and Science, project no. 2022/WK/14, and the National Science Center, contracts Opus 2021/41/B/ST2/01369 and 2021/43/B/ST2/01552 (Poland); the Funda\c{c}\~ao para a Ci\^encia e a Tecnologia, grant CEECIND/01334/2018 (Portugal); the National Priorities Research Program by Qatar National Research Fund; MCIN/AEI/10.13039/501100011033, ERDF ``a way of making Europe", and the Programa Estatal de Fomento de la Investigaci{\'o}n Cient{\'i}fica y T{\'e}cnica de Excelencia Mar\'{\i}a de Maeztu, grant MDM-2017-0765 and Programa Severo Ochoa del Principado de Asturias (Spain); the Chulalongkorn Academic into Its 2nd Century Project Advancement Project, and the National Science, Research and Innovation Fund via the Program Management Unit for Human Resources \& Institutional Development, Research and Innovation, grant B05F650021 (Thailand); the Kavli Foundation; the Nvidia Corporation; the SuperMicro Corporation; the Welch Foundation, contract C-1845; and the Weston Havens Foundation (USA).    
\end{acknowledgments}

\bibliography{auto_generated}

\providecommand{\href}[2]{#2}\begingroup\raggedright\begin{thebibliography}{10}%
\makeatletter
\providecommand{\hrefCMSnoop }[0]{\@secondoftwo}%
\makeatother
\providecommand{\doi}{\texttt{doi:}\begingroup \urlstyle{tt}\Url}

\bibitem{pdg}
\hrefCMSnoop {}{{Particle Data Group}, R.~L. Workman {et~al.}, ``Review of
  particle physics'',} \textit{ Prog. Theor. Exp. Phys.} \textbf{ 2022} (2022)
  083C01,
  \href{http://dx.doi.org/10.1093/ptep/ptac097}{\doi{10.1093/ptep/ptac097}}.
  Chapter 14 and references therein.

\bibitem{SNO:2002tuh}
\hrefCMSnoop {}{{SNO} Collaboration, ``Direct evidence for neutrino flavor
  transformation from neutral current interactions in the {Sudbury Neutrino
  Observatory}'',} \textit{ Phys. Rev. Lett.} \textbf{ 89} (2002) 011301,
  \href{http://dx.doi.org/10.1103/PhysRevLett.89.011301}{\doi{10.1103/PhysRevLett.89.011301}},
  \href{http://www.arXiv.org/abs/nucl-ex/0204008}{\texttt{arXiv:nucl-ex/0204008}}.

\bibitem{Super-Kamiokande:1998kpq}
\hrefCMSnoop {}{{Super-Kamiokande} Collaboration, ``Evidence for oscillation of
  atmospheric neutrinos'',} \textit{ Phys. Rev. Lett.} \textbf{ 81} (1998)
  1562,
  \href{http://dx.doi.org/10.1103/PhysRevLett.81.1562}{\doi{10.1103/PhysRevLett.81.1562}},
  \href{http://www.arXiv.org/abs/hep-ex/9807003}{\texttt{arXiv:hep-ex/9807003}}.

\bibitem{Super-Kamiokande:2004orf}
\hrefCMSnoop {}{{Super-Kamiokande} Collaboration, ``Evidence for an oscillatory
  signature in atmospheric neutrino oscillation'',} \textit{ Phys. Rev. Lett.}
  \textbf{ 93} (2004) 101801,
  \href{http://dx.doi.org/10.1103/PhysRevLett.93.101801}{\doi{10.1103/PhysRevLett.93.101801}},
  \href{http://www.arXiv.org/abs/hep-ex/0404034}{\texttt{arXiv:hep-ex/0404034}}.

\bibitem{PhysRevD.104.083504}
\hrefCMSnoop {}{E.~Di~Valentino, S.~Gariazzo, and O.~Mena, ``Most constraining
  cosmological neutrino mass bounds'',} \textit{ Phys. Rev. D} \textbf{ 104}
  (2021) 083504,
  \href{http://dx.doi.org/10.1103/PhysRevD.104.083504}{\doi{10.1103/PhysRevD.104.083504}},
  \href{http://www.arXiv.org/abs/2106.15267}{\texttt{arXiv:2106.15267}}.

\bibitem{KATRIN}
\hrefCMSnoop {}{{KATRIN} Collaboration, ``Direct neutrino-mass measurement with
  sub-electronvolt sensitivity'',} \textit{ Nature Phys.} \textbf{ 18} (2022)
  160,
  \href{http://dx.doi.org/10.1038/s41567-021-01463-1}{\doi{10.1038/s41567-021-01463-1}},
  \href{http://www.arXiv.org/abs/2105.08533}{\texttt{arXiv:2105.08533}}.

\bibitem{Maiezza:2010ic}
\hrefCMSnoop {}{A.~Maiezza, M.~Nemevsek, F.~Nesti, and G.~Senjanovic,
  ``Left-right symmetry at {LHC}'',} \textit{ Phys. Rev. D} \textbf{ 82} (2010)
  055022,
  \href{http://dx.doi.org/10.1103/PhysRevD.82.055022}{\doi{10.1103/PhysRevD.82.055022}},
\href{http://www.arXiv.org/abs/1005.5160}{\texttt{arXiv:1005.5160}}.

\bibitem{seesaw1}
\hrefCMSnoop {}{P.~Minkowski, ``$\mu \to \mathrm{e} \gamma$ at a rate of one
  out of 1-billion muon decays?'',} \textit{ Phys. Lett. B} \textbf{ 67} (1977)
  421,
\href{http://dx.doi.org/10.1016/0370-2693(77)90435-X}{\doi{10.1016/0370-2693(77)90435-X}}.

\bibitem{seesaw2}
\href {http://inspirehep.net/record/148634?ln=en}{M.~Gell-Mann, P.~Ramond, and
  R.~Slansky, ``Complex spinors and unified theories'',} in \textit{ in Proc.
  Supergravity Workshop at {S}tony {B}rook}, p.~341.
\newblock North-Holland, Amsterdam, Netherlands, 1979.

\bibitem{seesaw3}
\href {https://cds.cern.ch/record/120794}{T.~Yanagida, ``Horizontal gauge
  symmetry and masses of neutrinos'',} in \textit{ in Proc. Workshop on the
  Unified Theory and the Baryon Number in the Universe}, p.~95.
\newblock National Laboratory for High Energy Physics (KEK), 1979.

\bibitem{PhysRevLett.44.912}
\hrefCMSnoop {}{R.~N. Mohapatra and G.~Senjanovi{\'c}, ``Neutrino mass and
  spontaneous parity nonconservation'',} \textit{ Phys. Rev. Lett.} \textbf{
  44} (1980) 912,
  \href{http://dx.doi.org/10.1103/PhysRevLett.44.912}{\doi{10.1103/PhysRevLett.44.912}}.

\bibitem{DOI1981323}
M.~Doi\hrefCMSnoop {}{ {et~al.}, ``{CP} violation in {Majorana} neutrinos'',}
  \textit{ Phys. Lett. B} \textbf{ 102} (1981) 323,
  \href{http://dx.doi.org/10.1016/0370-2693(81)90627-4}{\doi{10.1016/0370-2693(81)90627-4}}.

\bibitem{CMS_Zp_dilep}
\hrefCMSnoop {}{{CMS Collaboration}, ``Search for resonant and nonresonant new
  phenomena in high-mass dilepton final states at $\sqrt{s}= 13$ {TeV}'',}
  \textit{ JHEP} \textbf{ 07} (2021) 208,
  \href{http://dx.doi.org/10.1007/JHEP07(2021)208}{\doi{10.1007/JHEP07(2021)208}},
  \href{http://www.arXiv.org/abs/2103.02708}{\texttt{arXiv:2103.02708}}.

\bibitem{CMS:2022yjm}
\hrefCMSnoop {}{{CMS Collaboration}, ``Search for new physics in the lepton
  plus missing transverse momentum final state in proton-proton collisions at
  $\sqrt{s}=13$ {TeV}'',} \textit{ JHEP} \textbf{ 07} (2022) 067,
  \href{http://dx.doi.org/10.1007/JHEP07(2022)067}{\doi{10.1007/JHEP07(2022)067}},
  \href{http://www.arXiv.org/abs/2202.06075}{\texttt{arXiv:2202.06075}}.

\bibitem{CMS_HNL_dilep}
\hrefCMSnoop {}{{CMS Collaboration}, ``Search for heavy majorana neutrinos in
  same-sign dilepton channels in proton-proton collisions at $\sqrt{s}=13$
  {TeV}'',} \textit{ JHEP} \textbf{ 01} (2019) 122,
  \href{http://dx.doi.org/10.1007/JHEP01(2019)122}{\doi{10.1007/JHEP01(2019)122}},
  \href{http://www.arXiv.org/abs/1806.10905}{\texttt{arXiv:1806.10905}}.

\bibitem{PhysRevLett.120.221801}
\hrefCMSnoop {}{{CMS Collaboration}, ``Search for heavy neutral leptons in
  events with three charged leptons in proton-proton collisions at
  $\sqrt{s}=13$ {TeV}'',} \textit{ Phys. Rev. Lett.} \textbf{ 120} (2018)
  221801,
  \href{http://dx.doi.org/10.1103/PhysRevLett.120.221801}{\doi{10.1103/PhysRevLett.120.221801}},
  \href{http://www.arXiv.org/abs/1802.02965}{\texttt{arXiv:1802.02965}}.

\bibitem{CMS:2022fut}
\hrefCMSnoop {}{{CMS Collaboration}, ``Search for long-lived heavy neutral
  leptons with displaced vertices in proton-proton collisions at $\sqrt{s}=13$
  {TeV}'',} \textit{ JHEP} \textbf{ 07} (2022) 081,
  \href{http://dx.doi.org/10.1007/JHEP07(2022)081}{\doi{10.1007/JHEP07(2022)081}},
  \href{http://www.arXiv.org/abs/2201.05578}{\texttt{arXiv:2201.05578}}.

\bibitem{ATLAS:2012ak}
\hrefCMSnoop {}{{ATLAS Collaboration}, ``Search for heavy neutrinos and
  right-handed {W} bosons in events with two leptons and jets in $pp$
  collisions at $\sqrt{s}=7$ {TeV} with the {ATLAS} detector'',} \textit{ Eur.
  Phys. J. C} \textbf{ 72} (2012) 2056,
  \href{http://dx.doi.org/10.1140/epjc/s10052-012-2056-4}{\doi{10.1140/epjc/s10052-012-2056-4}},
\href{http://www.arXiv.org/abs/1203.5420}{\texttt{arXiv:1203.5420}}.

\bibitem{CMS:2012zv}
\hrefCMSnoop {}{{CMS Collaboration}, ``Search for heavy neutrinos and ${W}_{R}$
  bosons with right-handed couplings in a left-right symmetric model in pp
  collisions at $\sqrt{s} = 7$ {TeV}'',} \textit{ Phys. Rev. Lett.} \textbf{
  109} (2012) 261802,
  \href{http://dx.doi.org/10.1103/PhysRevLett.109.261802}{\doi{10.1103/PhysRevLett.109.261802}},
\href{http://www.arXiv.org/abs/1210.2402}{\texttt{arXiv:1210.2402}}.

\bibitem{Khachatryan:2014dka}
\hrefCMSnoop {}{{CMS Collaboration}, ``Search for heavy neutrinos and $\mathrm
  {W}$ bosons with right-handed couplings in proton-proton collisions at
  $\sqrt{s} = 8$ {TeV}'',} \textit{ Eur. Phys. J. C} \textbf{ 74} (2014) 3149,
  \href{http://dx.doi.org/10.1140/epjc/s10052-014-3149-z}{\doi{10.1140/epjc/s10052-014-3149-z}},
\href{http://www.arXiv.org/abs/1407.3683}{\texttt{arXiv:1407.3683}}.

\bibitem{Aad:2015xaa}
\hrefCMSnoop {}{{ATLAS Collaboration}, ``Search for heavy majorana neutrinos
  with the {ATLAS} detector in pp collisions at $ \sqrt{s}=8 $ {TeV}'',}
  \textit{ JHEP} \textbf{ 07} (2015) 162,
  \href{http://dx.doi.org/10.1007/JHEP07(2015)162}{\doi{10.1007/JHEP07(2015)162}},
\href{http://www.arXiv.org/abs/1506.06020}{\texttt{arXiv:1506.06020}}.

\bibitem{WR_CMS_2015}
\hrefCMSnoop {}{{CMS Collaboration}, ``Search for heavy neutrinos or
  third-generation leptoquarks in final states with two hadronically decaying
  $\tau$ leptons and two jets in proton-proton collisions at $\sqrt{s}=13$
  {TeV}'',} \textit{ JHEP} \textbf{ 03} (2017) 077,
  \href{http://dx.doi.org/10.1007/JHEP03(2017)077}{\doi{10.1007/JHEP03(2017)077}},
  \href{http://www.arXiv.org/abs/1612.01190}{\texttt{arXiv:1612.01190}}.

\bibitem{WR_CMS_2016_ICHEP}
\hrefCMSnoop {}{{CMS Collaboration}, ``Search for third-generation scalar
  leptoquarks and heavy right-handed neutrinos in final states with two tau
  leptons and two jets in proton-proton collisions at $\sqrt{s}=13$ {TeV}'',}
  \textit{ JHEP} \textbf{ 07} (2017) 121,
  \href{http://dx.doi.org/10.1007/JHEP07(2017)121}{\doi{10.1007/JHEP07(2017)121}},
  \href{http://www.arXiv.org/abs/1703.03995}{\texttt{arXiv:1703.03995}}.

\bibitem{Sirunyan:2018pom}
\hrefCMSnoop {}{{CMS Collaboration}, ``Search for a heavy right-handed {W}
  boson and a heavy neutrino in events with two same-flavor leptons and two
  jets at $\sqrt{s}=13$ {TeV}'',} \textit{ JHEP} \textbf{ 05} (2018) 148,
  \href{http://dx.doi.org/10.1007/JHEP05(2018)148}{\doi{10.1007/JHEP05(2018)148}},
  \href{http://www.arXiv.org/abs/1803.11116}{\texttt{arXiv:1803.11116}}.

\bibitem{Aaboud:2018spl}
\hrefCMSnoop {}{{ATLAS Collaboration}, ``Search for heavy {Majorana} or {Dirac}
  neutrinos and right-handed {W} gauge bosons in final states with two charged
  leptons and two jets at $ \sqrt{s}=13 $ {TeV} with the {ATLAS} detector'',}
  \textit{ JHEP} \textbf{ 01} (2019) 016,
  \href{http://dx.doi.org/10.1007/JHEP01(2019)016}{\doi{10.1007/JHEP01(2019)016}},
\href{http://www.arXiv.org/abs/1809.11105}{\texttt{arXiv:1809.11105}}.

\bibitem{Aaboud:2019wfg}
\hrefCMSnoop {}{{ATLAS Collaboration}, ``Search for a right-handed gauge boson
  decaying into a high-momentum heavy neutrino and a charged lepton in $pp$
  collisions with the {ATLAS} detector at $\sqrt{s}=13$ {TeV}'',} \textit{
  Phys. Lett. B} \textbf{ 798} (2019) 134942,
  \href{http://dx.doi.org/10.1016/j.physletb.2019.134942}{\doi{10.1016/j.physletb.2019.134942}},
\href{http://www.arXiv.org/abs/1904.12679}{\texttt{arXiv:1904.12679}}.

\bibitem{CMS:2021dzb}
\hrefCMSnoop {}{{CMS Collaboration}, ``Search for a right-handed {W} boson and
  a heavy neutrino in proton-proton collisions at $\sqrt{s}= 13$ {TeV}'',}
  \textit{ JHEP} \textbf{ 04} (2022) 047,
  \href{http://dx.doi.org/10.1007/JHEP04(2022)047}{\doi{10.1007/JHEP04(2022)047}},
  \href{http://www.arXiv.org/abs/2112.03949}{\texttt{arXiv:2112.03949}}.

\bibitem{ATLAS:2023cjo}
\hrefCMSnoop {}{{ATLAS Collaboration}, ``Search for heavy majorana or dirac
  neutrinos and right-handed {W} gauge bosons in final states with charged
  leptons and jets in $pp$ collisions at $\sqrt{s}= 13$ {TeV} with the {ATLAS}
  detector'',} 2023.
  \href{http://www.arXiv.org/abs/2304.09553}{\texttt{arXiv:2304.09553}}.
Submitted to Eur. Phys. J. C.

\bibitem{hepdata}
\hrefCMSnoop {}{}{HEPD}ata record for this analysis, 2023.
\newblock
  \href{http://dx.doi.org/10.17182/hepdata.141323}{\doi{10.17182/hepdata.141323}}.

\bibitem{Chatrchyan:2008zzk}
\hrefCMSnoop {}{{CMS Collaboration}, ``The {CMS} experiment at the {CERN}
  {LHC}'',} \textit{ JINST} \textbf{ 3} (2008) S08004,
  \href{http://dx.doi.org/10.1088/1748-0221/3/08/S08004}{\doi{10.1088/1748-0221/3/08/S08004}}.

\bibitem{Alwall:2014hca}
J.~Alwall\hrefCMSnoop {}{ {et~al.}, ``The automated computation of tree-level
  and next-to-leading order differential cross sections, and their matching to
  parton shower simulations'',} \textit{ JHEP} \textbf{ 07} (2014) 079,
  \href{http://dx.doi.org/10.1007/JHEP07(2014)079}{\doi{10.1007/JHEP07(2014)079}},
\href{http://www.arXiv.org/abs/1405.0301}{\texttt{arXiv:1405.0301}}.

\bibitem{Alwall:2007fs}
J.~Alwall\hrefCMSnoop {}{ {et~al.}, ``Comparative study of various algorithms
  for the merging of parton showers and matrix elements in hadronic
  collisions'',} \textit{ Eur. Phys. J. C} \textbf{ 53} (2008) 473,
  \href{http://dx.doi.org/10.1140/epjc/s10052-007-0490-5}{\doi{10.1140/epjc/s10052-007-0490-5}},
\href{http://www.arXiv.org/abs/0706.2569}{\texttt{arXiv:0706.2569}}.

\bibitem{Nason:2004rx}
\hrefCMSnoop {}{P.~Nason, ``{A new method for combining NLO QCD with shower
  Monte Carlo algorithms}'',} \textit{ JHEP} \textbf{ 11} (2004) 040,
  \href{http://dx.doi.org/10.1088/1126-6708/2004/11/040}{\doi{10.1088/1126-6708/2004/11/040}},
\href{http://www.arXiv.org/abs/hep-ph/0409146}{\texttt{arXiv:hep-ph/0409146}}.

\bibitem{Frixione:2007vw}
\hrefCMSnoop {}{S.~Frixione, P.~Nason, and C.~Oleari, ``{Matching NLO QCD
  computations with parton shower simulations: the \POWHEG method}'',} \textit{
  JHEP} \textbf{ 11} (2007) 070,
  \href{http://dx.doi.org/10.1088/1126-6708/2007/11/070}{\doi{10.1088/1126-6708/2007/11/070}},
\href{http://www.arXiv.org/abs/0709.2092}{\texttt{arXiv:0709.2092}}.

\bibitem{Alioli:2010xd}
\hrefCMSnoop {}{S.~Alioli, P.~Nason, C.~Oleari, and E.~Re, ``{A general
  framework for implementing NLO calculations in shower Monte Carlo programs:
  the \POWHEG BOX}'',} \textit{ JHEP} \textbf{ 06} (2010) 043,
  \href{http://dx.doi.org/10.1007/JHEP06(2010)043}{\doi{10.1007/JHEP06(2010)043}},
\href{http://www.arXiv.org/abs/1002.2581}{\texttt{arXiv:1002.2581}}.

\bibitem{Frixione:2007nw}
\hrefCMSnoop {}{S.~Frixione, P.~Nason, and G.~Ridolfi, ``{A positive-weight
  next-to-leading-order Monte Carlo for heavy flavour hadroproduction}'',}
  \textit{ JHEP} \textbf{ 09} (2007) 126,
  \href{http://dx.doi.org/10.1088/1126-6708/2007/09/126}{\doi{10.1088/1126-6708/2007/09/126}},
  \href{http://www.arXiv.org/abs/0707.3088}{\texttt{arXiv:0707.3088}}.

\bibitem{Sjostrand:2014zea}
T.~Sj{\"o}strand\hrefCMSnoop {}{ {et~al.}, ``{An introduction to PYTHIA
  8.2}'',} \textit{ Comput. Phys. Commun.} \textbf{ 191} (2015) 159,
  \href{http://dx.doi.org/10.1016/j.cpc.2015.01.024}{\doi{10.1016/j.cpc.2015.01.024}},
\href{http://www.arXiv.org/abs/1410.3012}{\texttt{arXiv:1410.3012}}.

\bibitem{Alioli:2009je}
\hrefCMSnoop {}{S.~Alioli, P.~Nason, C.~Oleari, and E.~Re, ``{NLO single-top
  production matched with shower in POWHEG: s- and t-channel contributions}'',}
  \textit{ JHEP} \textbf{ 09} (2009) 111,
  \href{http://dx.doi.org/10.1088/1126-6708/2009/09/111}{\doi{10.1088/1126-6708/2009/09/111}},
  \href{http://www.arXiv.org/abs/0907.4076}{\texttt{arXiv:0907.4076}}.
  [Erratum: JHEP 02, 011 (2010)].

\bibitem{Re:2010bp}
\hrefCMSnoop {}{E.~Re, ``{Single-top Wt-channel production matched with parton
  showers using the POWHEG method}'',} \textit{ Eur. Phys. J. C} \textbf{ 71}
  (2011) 1547,
  \href{http://dx.doi.org/10.1140/epjc/s10052-011-1547-z}{\doi{10.1140/epjc/s10052-011-1547-z}},
  \href{http://www.arXiv.org/abs/1009.2450}{\texttt{arXiv:1009.2450}}.

\bibitem{Ball:2014uwa}
\hrefCMSnoop {}{{NNPDF} Collaboration, ``Parton distributions for the {LHC}
  {Run} {II}'',} \textit{ JHEP} \textbf{ 04} (2015) 040,
  \href{http://dx.doi.org/10.1007/JHEP04(2015)040}{\doi{10.1007/JHEP04(2015)040}},
\href{http://www.arXiv.org/abs/1410.8849}{\texttt{arXiv:1410.8849}}.

\bibitem{Ball:2017nwa}
\hrefCMSnoop {}{{NNPDF} Collaboration, ``Parton distributions from
  high-precision collider data'',} \textit{ Eur. Phys. J. C} \textbf{ 77}
  (2017) 663,
  \href{http://dx.doi.org/10.1140/epjc/s10052-017-5199-5}{\doi{10.1140/epjc/s10052-017-5199-5}},
\href{http://www.arXiv.org/abs/1706.00428}{\texttt{arXiv:1706.00428}}.

\bibitem{Khachatryan:2015pea}
\hrefCMSnoop {}{{CMS Collaboration}, ``Event generator tunes obtained from
  underlying event and multiparton scattering measurements'',} \textit{ Eur.
  Phys. J. C} \textbf{ 76} (2016) 155,
  \href{http://dx.doi.org/10.1140/epjc/s10052-016-3988-x}{\doi{10.1140/epjc/s10052-016-3988-x}},
  \href{http://www.arXiv.org/abs/1512.00815}{\texttt{arXiv:1512.00815}}.

\bibitem{Sirunyan:2019dfx}
\hrefCMSnoop {}{{CMS Collaboration}, ``Extraction and validation of a new set
  of {CMS} {PYTHIA8} tunes from underlying-event measurements'',} \textit{ Eur.
  Phys. J. C} \textbf{ 80} (2020) 4,
  \href{http://dx.doi.org/10.1140/epjc/s10052-019-7499-4}{\doi{10.1140/epjc/s10052-019-7499-4}},
\href{http://www.arXiv.org/abs/1903.12179}{\texttt{arXiv:1903.12179}}.

\bibitem{Geant4}
\hrefCMSnoop {}{{GEANT4} Collaboration, ``{\GEANTfour}---a simulation
  toolkit'',} \textit{ Nucl. Instrum. Meth. A} \textbf{ 506} (2003) 250,
\href{http://dx.doi.org/10.1016/S0168-9002(03)01368-8}{\doi{10.1016/S0168-9002(03)01368-8}}.

\bibitem{Mattelaer:2016ynf}
\hrefCMSnoop {}{O.~Mattelaer, M.~Mitra, and R.~Ruiz, ``Automated neutrino jet
  and top jet predictions at next-to-leading-order with parton shower matching
  in effective left-right symmetric models'',} 2016.
  \href{http://www.arXiv.org/abs/1610.08985}{\texttt{arXiv:1610.08985}}.

\bibitem{Khachatryan:2016bia}
\hrefCMSnoop {}{{CMS Collaboration}, ``{The CMS trigger system}'',} \textit{
  JINST} \textbf{ 12} (2017) P01020,
  \href{http://dx.doi.org/10.1088/1748-0221/12/01/P01020}{\doi{10.1088/1748-0221/12/01/P01020}},
\href{http://www.arXiv.org/abs/1609.02366}{\texttt{arXiv:1609.02366}}.

\bibitem{CMS-TDR-15-02}
\href {http://cds.cern.ch/record/2020886}{{CMS Collaboration}, ``Technical
  proposal for the {Phase-II} upgrade of the {Compact Muon Solenoid}'',} CMS
  Technical Proposal CERN-LHCC-2015-010, CMS-TDR-15-02, 2015.

\bibitem{CMS-PRF-14-001}
\hrefCMSnoop {}{{CMS Collaboration}, ``Particle-flow reconstruction and global
  event description with the {CMS} detector'',} \textit{ JINST} \textbf{ 12}
  (2017) P10003,
  \href{http://dx.doi.org/10.1088/1748-0221/12/10/P10003}{\doi{10.1088/1748-0221/12/10/P10003}},
\href{http://www.arXiv.org/abs/1706.04965}{\texttt{arXiv:1706.04965}}.

\bibitem{CMS:2020uim}
\hrefCMSnoop {}{{CMS Collaboration}, ``Electron and photon reconstruction and
  identification with the {CMS} experiment at the {CERN} {LHC}'',} \textit{
  JINST} \textbf{ 16} (2021) P05014,
  \href{http://dx.doi.org/10.1088/1748-0221/16/05/P05014}{\doi{10.1088/1748-0221/16/05/P05014}},
  \href{http://www.arXiv.org/abs/2012.06888}{\texttt{arXiv:2012.06888}}.

\bibitem{CMS-DP-2020-021}
\href {https://cds.cern.ch/record/2717925}{{CMS Collaboration}, ``{ECAL} 2016
  refined calibration and {Run2} summary plots'',} CMS Detector Performance
  Note CMS-DP-2020-021, 2020.

\bibitem{Sirunyan_2018}
\hrefCMSnoop {}{{CMS Collaboration}, ``Performance of the {CMS} muon detector
  and muon reconstruction with proton-proton collisions at $\sqrt{s}=$ 13
  {TeV}'',} \textit{ JINST} \textbf{ 13} (2018) P06015,
  \href{http://dx.doi.org/10.1088/1748-0221/13/06/p06015}{\doi{10.1088/1748-0221/13/06/p06015}},
  \href{http://www.arXiv.org/abs/1804.04528}{\texttt{arXiv:1804.04528}}.

\bibitem{Sirunyan:2018exx}
\hrefCMSnoop {}{{CMS Collaboration}, ``{Search for high-mass resonances in
  dilepton final states in proton-proton collisions at $\sqrt{s}=$ 13 TeV}'',}
  \textit{ JHEP} \textbf{ 06} (2018) 120,
  \href{http://dx.doi.org/10.1007/JHEP06(2018)120}{\doi{10.1007/JHEP06(2018)120}},
\href{http://www.arXiv.org/abs/1803.06292}{\texttt{arXiv:1803.06292}}.

\bibitem{Cacciari:2008gp}
\hrefCMSnoop {}{M.~Cacciari, G.~P. Salam, and G.~Soyez, ``{The anti-\kt jet
  clustering algorithm}'',} \textit{ JHEP} \textbf{ 04} (2008) 063,
  \href{http://dx.doi.org/10.1088/1126-6708/2008/04/063}{\doi{10.1088/1126-6708/2008/04/063}},
  \href{http://www.arXiv.org/abs/0802.1189}{\texttt{arXiv:0802.1189}}.

\bibitem{Cacciari:2011ma}
\hrefCMSnoop {}{M.~Cacciari, G.~P. Salam, and G.~Soyez, ``{FastJet user
  manual}'',} \textit{ Eur. Phys. J. C} \textbf{ 72} (2012) 1896,
  \href{http://dx.doi.org/10.1140/epjc/s10052-012-1896-2}{\doi{10.1140/epjc/s10052-012-1896-2}},
\href{http://www.arXiv.org/abs/1111.6097}{\texttt{arXiv:1111.6097}}.

\bibitem{Bertolini:2014bba}
\hrefCMSnoop {}{D.~Bertolini, P.~Harris, M.~Low, and N.~Tran, ``Pileup per
  particle identification'',} \textit{ JHEP} \textbf{ 10} (2014) 059,
  \href{http://dx.doi.org/10.1007/JHEP10(2014)059}{\doi{10.1007/JHEP10(2014)059}},
\href{http://www.arXiv.org/abs/1407.6013}{\texttt{arXiv:1407.6013}}.

\bibitem{CMS:2020ebo}
\hrefCMSnoop {}{{CMS Collaboration}, ``Pileup mitigation at {CMS} in 13 {\TeV}
  data'',} \textit{ JINST} \textbf{ 15} (2020) P09018,
  \href{http://dx.doi.org/10.1088/1748-0221/15/09/P09018}{\doi{10.1088/1748-0221/15/09/P09018}},
  \href{http://www.arXiv.org/abs/2003.00503}{\texttt{arXiv:2003.00503}}.

\bibitem{CMS-PAS-JME-16-003}
\href {https://cds.cern.ch/record/2256875}{{CMS Collaboration}, ``{Jet
  algorithms performance in 13{\TeV} data}'',} {CMS Physics Analysis Summary}
  CMS-PAS-JME-16-003, 2017.

\bibitem{Dasgupta:2013ihk}
\hrefCMSnoop {}{M.~Dasgupta, A.~Fregoso, S.~Marzani, and G.~P. Salam, ``Towards
  an understanding of jet substructure'',} \textit{ JHEP} \textbf{ 09} (2013)
  029,
  \href{http://dx.doi.org/10.1007/JHEP09(2013)029}{\doi{10.1007/JHEP09(2013)029}},
\href{http://www.arXiv.org/abs/1307.0007}{\texttt{arXiv:1307.0007}}.

\bibitem{Butterworth:2008iy}
\hrefCMSnoop {}{J.~M. Butterworth, A.~R. Davison, M.~Rubin, and G.~P. Salam,
  ``Jet substructure as a new {Higgs} search channel at the {LHC}'',} \textit{
  Phys. Rev. Lett.} \textbf{ 100} (2008) 242001,
  \href{http://dx.doi.org/10.1103/PhysRevLett.100.242001}{\doi{10.1103/PhysRevLett.100.242001}},
\href{http://www.arXiv.org/abs/0802.2470}{\texttt{arXiv:0802.2470}}.

\bibitem{Dokshitzer:1997in}
\hrefCMSnoop {}{Y.~L. Dokshitzer, G.~D. Leder, S.~Moretti, and B.~R. Webber,
  ``Better jet clustering algorithms'',} \textit{ JHEP} \textbf{ 08} (1997)
  001,
  \href{http://dx.doi.org/10.1088/1126-6708/1997/08/001}{\doi{10.1088/1126-6708/1997/08/001}},
\href{http://www.arXiv.org/abs/hep-ph/9707323}{\texttt{arXiv:hep-ph/9707323}}.

\bibitem{Wobisch:1998wt}
\href {https://inspirehep.net/record/484872}{M.~Wobisch and T.~Wengler,
  ``Hadronization corrections to jet cross-sections in deep inelastic
  scattering'',} in \textit{ {Workshop on Monte Carlo Generators for HERA
  Physics, Hamburg, Germany}}, p.~270.
\newblock 1998.
\newblock
\href{http://www.arXiv.org/abs/hep-ph/9907280}{\texttt{arXiv:hep-ph/9907280}}.
\newblock

\bibitem{Larkoski:2014wba}
\hrefCMSnoop {}{A.~J. Larkoski, S.~Marzani, G.~Soyez, and J.~Thaler, ``Soft
  drop'',} \textit{ JHEP} \textbf{ 05} (2014) 146,
  \href{http://dx.doi.org/10.1007/JHEP05(2014)146}{\doi{10.1007/JHEP05(2014)146}},
\href{http://www.arXiv.org/abs/1402.2657}{\texttt{arXiv:1402.2657}}.

\bibitem{Czakon:2011xx}
\hrefCMSnoop {}{M.~Czakon and A.~Mitov, ``Top++: A program for the calculation
  of the top-pair cross-section at hadron colliders'',} \textit{ Comput. Phys.
  Commun.} \textbf{ 185} (2014) 2930,
  \href{http://dx.doi.org/10.1016/j.cpc.2014.06.021}{\doi{10.1016/j.cpc.2014.06.021}},
  \href{http://www.arXiv.org/abs/1112.5675}{\texttt{arXiv:1112.5675}}.

\bibitem{Sirunyan:2019bzr}
\hrefCMSnoop {}{{CMS Collaboration}, ``Measurements of differential {Z} boson
  production cross sections in proton-proton collisions at $ \sqrt{s} $ = 13
  {TeV}'',} \textit{ JHEP} \textbf{ 12} (2019) 061,
  \href{http://dx.doi.org/10.1007/JHEP12(2019)061}{\doi{10.1007/JHEP12(2019)061}},
\href{http://www.arXiv.org/abs/1909.04133}{\texttt{arXiv:1909.04133}}.

\bibitem{cms_jet_performance_8TeV}
\hrefCMSnoop {}{{CMS Collaboration}, ``{Jet energy scale and resolution in the
  CMS experiment in pp collisions at 8 TeV}'',} \textit{ JINST} \textbf{ 12}
  (2017) P02014,
  \href{http://dx.doi.org/10.1088/1748-0221/12/02/P02014}{\doi{10.1088/1748-0221/12/02/P02014}},
\href{http://www.arXiv.org/abs/1607.03663}{\texttt{arXiv:1607.03663}}.

\bibitem{CMS-LUM-17-003}
\hrefCMSnoop {}{{CMS Collaboration}, ``Precision luminosity measurement in
  proton-proton collisions at $\sqrt{s} =$ 13 {TeV} in 2015 and 2016 at
  {CMS}'',} \textit{ Eur. Phys. J. C} \textbf{ 81} (2021) 800,
  \href{http://dx.doi.org/10.1140/epjc/s10052-021-09538-2}{\doi{10.1140/epjc/s10052-021-09538-2}},
  \href{http://www.arXiv.org/abs/2104.01927}{\texttt{arXiv:2104.01927}}.

\bibitem{CMS-PAS-LUM-17-004}
\href {https://cds.cern.ch/record/2621960/}{{CMS Collaboration}, ``{CMS}
  luminosity measurement for the 2017 data-taking period at $\sqrt{s}$ = 13
  {TeV}'',} CMS Physics Analysis Summary CMS-PAS-LUM-17-004, 2018.

\bibitem{CMS-PAS-LUM-18-002}
\href {https://cds.cern.ch/record/2676164/}{{CMS Collaboration}, ``{CMS}
  luminosity measurement for the 2018 data-taking period at $\sqrt{s}$ = 13
  {TeV}'',} CMS Physics Analysis Summary CMS-PAS-LUM-18-002, 2019.

\bibitem{cms_level1_trigger_performance}
\hrefCMSnoop {}{{CMS Collaboration}, ``Performance of the {CMS} {Level-1}
  trigger in proton-proton collisions at $\sqrt{s}$ = 13 {TeV}'',} \textit{
  JINST} \textbf{ 15} (2020) P10017,
  \href{http://dx.doi.org/10.1088/1748-0221/15/10/p10017}{\doi{10.1088/1748-0221/15/10/p10017}},
  \href{http://www.arXiv.org/abs/2006.10165}{\texttt{arXiv:2006.10165}}.

\bibitem{Butterworth_2016}
\hrefCMSnoop {}{J.~Butterworth {et~al.}, ``{PDF4LHC} recommendations for {LHC}
  {Run} {II}'',} \textit{ J. Phys. G} \textbf{ 43} (2016) 023001,
  \href{http://dx.doi.org/10.1088/0954-3899/43/2/023001}{\doi{10.1088/0954-3899/43/2/023001}},
  \href{http://www.arXiv.org/abs/1510.03865}{\texttt{arXiv:1510.03865}}.

\bibitem{CMS:2016mku}
\hrefCMSnoop {}{{CMS Collaboration}, ``Search for new physics in same-sign
  dilepton events in proton-proton collisions at $\sqrt{s} = 13$ {TeV}'',}
  \textit{ Eur. Phys. J. C} \textbf{ 76} (2016) 439,
  \href{http://dx.doi.org/10.1140/epjc/s10052-016-4261-z}{\doi{10.1140/epjc/s10052-016-4261-z}},
  \href{http://www.arXiv.org/abs/1605.03171}{\texttt{arXiv:1605.03171}}.

\bibitem{Cowan:2010js}
\hrefCMSnoop {}{G.~Cowan, K.~Cranmer, E.~Gross, and O.~Vitells, ``Asymptotic
  formulae for likelihood-based tests of new physics'',} \textit{ Eur. Phys. J.
  C} \textbf{ 71} (2011) 1554,
  \href{http://dx.doi.org/10.1140/epjc/s10052-011-1554-0}{\doi{10.1140/epjc/s10052-011-1554-0}},
  \href{http://www.arXiv.org/abs/1007.1727}{\texttt{arXiv:1007.1727}}.
[Erratum: \DOI{10.1140/epjc/s10052-013-2501-z}].

\bibitem{Junk:1999kv}
\hrefCMSnoop {}{T.~Junk, ``{Confidence level computation for combining searches
  with small statistics}'',} \textit{ Nucl. Instrum. Meth. A} \textbf{ 434}
  (1999) 435,
  \href{http://dx.doi.org/10.1016/S0168-9002(99)00498-2}{\doi{10.1016/S0168-9002(99)00498-2}},
\href{http://www.arXiv.org/abs/hep-ex/9902006}{\texttt{arXiv:hep-ex/9902006}}.

\bibitem{Read:2002hq}
\hrefCMSnoop {}{A.~L. Read, ``Presentation of search results: {the $\CLs$
  technique}'',} \textit{ J. Phys. G} \textbf{ 28} (2002) 2693,
\href{http://dx.doi.org/10.1088/0954-3899/28/10/313}{\doi{10.1088/0954-3899/28/10/313}}.

\bibitem{CMS:2019gwf}
\hrefCMSnoop {}{{CMS Collaboration}, ``Search for high mass dijet resonances
  with a new background prediction method in proton-proton collisions at
  $\sqrt{s} =$ 13 {TeV}'',} \textit{ JHEP} \textbf{ 05} (2020) 033,
  \href{http://dx.doi.org/10.1007/JHEP05(2020)033}{\doi{10.1007/JHEP05(2020)033}},
  \href{http://www.arXiv.org/abs/1911.03947}{\texttt{arXiv:1911.03947}}.

\end{thebibliography}\endgroup
\cleardoublepage \appendix\section{The CMS Collaboration \label{app:collab}}\begin{sloppypar}\hyphenpenalty=5000\widowpenalty=500\clubpenalty=5000
\cmsinstitute{Yerevan Physics Institute, Yerevan, Armenia}
{\tolerance=6000
A.~Tumasyan\cmsorcid{0009-0000-0684-6742}
\par}
\cmsinstitute{Institut f\"{u}r Hochenergiephysik, Vienna, Austria}
{\tolerance=6000
W.~Adam\cmsorcid{0000-0001-9099-4341}, J.W.~Andrejkovic, T.~Bergauer\cmsorcid{0000-0002-5786-0293}, S.~Chatterjee\cmsorcid{0000-0003-2660-0349}, K.~Damanakis\cmsorcid{0000-0001-5389-2872}, M.~Dragicevic\cmsorcid{0000-0003-1967-6783}, A.~Escalante~Del~Valle\cmsorcid{0000-0002-9702-6359}, P.S.~Hussain\cmsorcid{0000-0002-4825-5278}, M.~Jeitler\cmsAuthorMark{1}\cmsorcid{0000-0002-5141-9560}, N.~Krammer\cmsorcid{0000-0002-0548-0985}, L.~Lechner\cmsorcid{0000-0002-3065-1141}, D.~Liko\cmsorcid{0000-0002-3380-473X}, I.~Mikulec\cmsorcid{0000-0003-0385-2746}, P.~Paulitsch, F.M.~Pitters, J.~Schieck\cmsAuthorMark{1}\cmsorcid{0000-0002-1058-8093}, R.~Sch\"{o}fbeck\cmsorcid{0000-0002-2332-8784}, D.~Schwarz\cmsorcid{0000-0002-3821-7331}, S.~Templ\cmsorcid{0000-0003-3137-5692}, W.~Waltenberger\cmsorcid{0000-0002-6215-7228}, C.-E.~Wulz\cmsAuthorMark{1}\cmsorcid{0000-0001-9226-5812}
% ...
\par}
\vskip\cmsinstskip
\dag:~Deceased\\
$^{1}$Also at TU Wien, Vienna, Austria\\
% ...
\end{sloppypar}
\end{document}